\documentclass[12pt,a4]{article}

\usepackage{amsmath}
\usepackage[dvips]{graphicx}

\def\D{\nabla}
\def\oD{\overline{\D}}
\def\U{\mathcal{U}}

\def\Dc{\mathcal{D}}
\def\olambda{\overline{\lambda}}
\def\orho{\overline{\rho}}
\def\oLambda{\overline{\Lambda}}
\def\oUpsilon{\overline{\Upsilon}}
\def\oD{\overline{\D}}
\def\oG{\overline{G}}
\def\cG{\mathcal{G}}
\def\ocG{\overline{\mathcal{G}}}
\def\cK{\mathcal{K}}
\def\os{\overline{s}}

\makeatletter
\@addtoreset{equation}{section}
\makeatother

\begin{document}

\begin{flushright}
October, 2007
\end{flushright}

\vspace{0.0cm}

\begin{center}

{\Large \bf On the Continuum and Lattice Formulations of \\
\vspace{0.3cm}

$N=4\ D=3$ Twisted Super Yang-Mills }

\vspace{1cm}

{ Kazuhiro Nagata}\footnote{knagata@indiana.edu}
\\

\vspace{0.5cm}
{\it{ Department of Physics, Indiana University }}\\
{\it{ Bloomington, IN 47405, USA}}\\
\end{center}

\vspace{0.5cm}

\begin{abstract}
Employing a twisted superspace with eight supercharges,
we describe an off-shell formulation of $N=4\ D=3$ twisted
super Yang-Mills in the continuum spacetime
which underlies the recent proposal of
$N=4\ D=3$ twisted super Yang-Mills on a lattice \cite{DKKN3}.
By a dimensional reduction from the $N=2\ D=4$,
we explore the two possible topological twists of $N=4\ D=3$
and then show that the lattice formulation given in \cite{DKKN3}
is essentially categorized as the B-type. 
We also show that,
amongst the two inequivalent 
twists of $N=4\ D=3$, 
only the B-type SYM can be realized on the lattice 
consistently with the Leibniz rule 
and the gauge covariance on the lattice.
\end{abstract}

\section{Introduction}
It has been two decades since the twisted supersymmetry (SUSY)
was first introduced \cite{Witten1}.
Although 
the notion of twisted SUSY was originally proposed
in the context of topological field theories \cite{Witten1,Witten2,Witten3},
it has been recently paid much attention also 
from the lattice SUSY point of view
\cite{DKKN1,DKKN2,DKKN3,twist-lattsusy}.
The main purpose of lattice SUSY is to provide
a constructive formulation of supersymmetric models.
To this end there have been a wide variety of studies 
addressing this subject 
\cite{old-lattsusy,D-W-lattsusy,deconstruction,other-lattsusy1}. 
On the other hand, it has long been 
recognized that there are a couple of
obstacles in formulating lattice SUSY,
such as the breakdown of Leibniz rule
and the existence of fermion doubling on the lattice.
The twisted SUSY is providing a fundamental framework 
to overcome these difficulties.
In the series of studies \cite{DKKN1,DKKN2,DKKN3},
we formulated the $N=D=2$ twisted Wess-Zumino type models,
$N=D=2$ twisted super Yang-Mills (SYM) and $N=4\ D=3$
twisted SYM on a lattice.
Starting from a careful observation of difference operators 
and supercharges, we introduced the notion of lattice Leibniz rule conditions.
We then explicitly showed that the Dirac-K\"ahler twisted SUSY algebra can satisfy 
these conditions.
As a result, we could realize all the supercharges 
and the invariance on the lattice
by systematically introducing 
the link (anti-)commutator algebra.
It is crucial to observe that
the importance of the twisted SUSY can be traced back to the
intrinsic relation between twisted fermions and Dirac-K\"ahler fermions
\cite{KT,KKU,KKM}.
Furthermore, the very recent development in terms of the matrix
formulation \cite{ADKS} is serving as a fundamental framework
realizing the above picture more rigidly. 
It is also interesting to mention that 
the above link approach of lattice SUSY and
the so-called orbifold approach of lattice SUSY \cite{deconstruction} turn to be 
closely related each other, which was mentioned in the discussion in \cite{DKKN2}
and also recently pointed out in \cite{link_deconst}.

Keeping these circumstances in mind, in this paper, 
we describe a manifestly off-shell formulation of
the $N=4\ D=3$ twisted SYM in the 
continuum flat spacetime which underlies our recent proposal of
the $N=4\ D=3$ twisted SYM on the lattice.
Studies of the $N=4\ D=3$ twisted SYM in the continuum spacetime
have been given in the past \cite{BT,GM} with the classification of
two inequivalent topological twists which are called the super BF type (A-type)
and the Blau-Thompson type (B-type).
The twisted SYM multiplets and the algebra 
given in these studies are not entirely off-shell.
In this paper,
we explore these inequivalent twists
entirely in a off-shell regime.
We then show that the lattice SYM multiplet given in \cite{DKKN3}
is categorized as the B-type twist. 
We also investigate these two inequivalent twists from the lattice point of view
and show that 
only the B-type multiplet can be realized on the lattice 
consistently with the Leibniz rule and gauge covariance on the lattice.

This paper is organized as follows.
In Sec. 2, starting from the $N=4\ D=3$ SUSY algebra,
we introduce a twisted $N=4\ D=3$ superspace formulation.
We employ an extended SUSY superconnection method
in order to provide a manifest gauge covariant off-shell framework.
Introducing the $N=4\ D=3$ twisted SYM constraints
for the supercovariant derivatives,
we construct a manifestly invariant formulation of $N=4\ D=3$ twisted SYM
in the continuum spacetime which underlies the recent proposal of
twisted SYM on a three dimensional lattice \cite{DKKN3}.
We also discuss about the twisted SUSY exact
relation between the super Chern-Simons and the SYM
completely in the off-shell regime,
stressing that the existence of sub-algebra
and sub-multiplet is responsible for the off-shell super
Chern-Simons realization.
In Sec. 3, by a dimensional reduction from the $N=2\ D=4$ twisted SYM, 
we explore the two possible twists of $N=4\ D=3$.
We then show that the lattice formulation 
of SYM given in \cite{DKKN3} is classified as the B-type twisted SYM.
In Sec. 4, after reviewing the notion of the lattice Leibniz rule
\cite{DKKN1,DKKN2,DKKN3},
we examine the possibilities of realizing the twisted $N=2\ D=4$,
$N=4\ D=3$ A-type and B-type SYM on the lattice. 
We then explicitly show that only the B-type
twisted SYM can be consistent with the lattice Leibniz rule and the
gauge covariance on the lattice.
Namely, the formulation given in \cite{DKKN3}
is the unique lattice realization of $N=4\ D=3$ twisted SYM satisfying 
these conditions.
Sec. 5 gives the summary and the discussions.


\section{Superspace formulation of $N=4\ D=3$ twisted SYM}
\label{offshellSYM}

In this section, employing a twisted superfield method with eight supercharges,
we explicitly perform an off-shell construction of 
the $N=4\ D=3$ twisted SYM.
It provides the underlying continuum theory for the recent proposal of
twisted SYM on the three dimensional lattice \cite{DKKN3}.
We start from the following $N=4\ D=3$ 
SUSY algebra, 
\begin{eqnarray}
&&\hspace{20pt} \{Q_{\alpha i},\overline{Q}_{j\beta}\} \ =\ 
2\delta_{ij}(\gamma_{\mu})_{\alpha\beta}P_{\mu}, \label{N=4D=3SUSYalgebra}\\[2pt]
[J_{\mu},Q_{\alpha i}] &=& +\frac{1}{2}(\gamma_{\mu})_{\alpha\beta}Q_{\beta i},
\hspace{30pt}
[J_{\mu},\overline{Q}_{i\alpha}] \ =\ -\frac{1}{2}
\overline{Q}_{i\beta}(\gamma_{\mu})_{\beta\alpha},\\[0pt]
[R_{\mu},Q_{\alpha i}] &=& -\frac{1}{2}Q_{\alpha j}(\gamma_{\mu})_{ji},
\hspace{30pt}
[R_{\mu},\overline{Q}_{i\alpha}] \ =\ +\frac{1}{2}
(\gamma_{\mu})_{ij}\overline{Q}_{j\alpha},\\[2pt]
[J_{\mu},P_{\nu}] &=& -i\epsilon_{\mu\nu\rho}P_{\rho},
\hspace{20pt}
[J_{\mu},J_{\nu}] \ =\ -i\epsilon_{\mu\nu\rho}J_{\rho},
\hspace{20pt}
[R_{\mu},R_{\nu}] \ =\ -i\epsilon_{\mu\nu\rho}R_{\rho}, \\[4pt]
[R_{\mu},P_{\nu}] &=&
[P_{\mu},P_{\nu}] \ =\ 
[J_{\mu},R_{\nu}] \ =\ 0,
\end{eqnarray}
where the gamma matrices $\gamma_{\mu}$
can be taken as the Pauli matrices,
$\gamma^{\mu}(\mu=1,2,3)\equiv (\sigma^{1},\sigma^{2},\sigma^{3})$.
The conjugate supercharge $\overline{Q}_{i\alpha}$ 
can be taken as the complex conjugation
of $Q_{\alpha i}$, $\overline{Q}_{i\alpha}=Q^{*}_{\alpha i}$.
The $J_{\mu}$ and $R_{\mu}\ (\mu =1,2,3)$ are the generators of 
$SO(3)_{E}\simeq SU(2)_{E}$ Euclidean Lorentz rotations 
and $SO(3)_{R}\simeq SU(2)_{R}$ internal rotations, respectively.

As in the case of $N=D=2$ \cite{KT, KKU, N=D=2twist}
and $N=D=4$ \cite{KKM,BT,N=4twist}, 
the twisting procedure can be performed
by taking the diagonal subgroup of 
the Lorentz rotations and the internal rotations.
Here in the case of $N=4\ D=3$, we take the diagonal subgroup
$(SO(3)_{E}\times SO(3)_{R})_{diag}$ 
whose covering group is $(SU(2)_{E}\times SU(2)_{R})_{diag}$.
This corresponds to 
introducing the twisted Lorentz generator $J^{diag}_{\mu}$ as
a diagonal sum of $J_{\mu}$ and $R_{\mu}$,
$J^{diag}_{\mu} \equiv J_{\mu} + R_{\mu}$.
Since, after the twisting, the Lorentz index $\alpha$ and internal index $i$
are rotated on the same footing,
the resulting algebra is most naturally expressed in terms of the following
Dirac-K\"ahler expansion of the supercharges,
\begin{eqnarray}
Q_{\alpha i} &=& (\mathbf{1}s+\gamma_{\mu}s_{\mu})_{\alpha i}, 
\hspace{30pt}
\overline{Q}_{i\alpha} \ =\ 
(\mathbf{1}\overline{s}+\gamma_{\mu}\overline{s}_{\mu})_{i\alpha},
\end{eqnarray}
where $\bf{1}$ represents the two-by-two unit matrix.
The coefficients 
$(s,\overline{s}_{\mu},s_{\mu},\overline{s})$
are called the $N=4\ D=3$ twisted supercharges.
After the expansions, the original SUSY algebra (\ref{N=4D=3SUSYalgebra})
can be expressed as,
\begin{eqnarray}
\{s,\overline{s}_{\mu}\} &=& P_{\mu}, \label{N=4D=3twistedSUSYalgebra1}\\[2pt]
\{s_{\mu},\overline{s}_{\nu}\} & =& -i\epsilon_{\mu\nu\rho}P_{\rho},
\label{N=4D=3twistedSUSYalgebra2}\\[2pt]
\{\overline{s},s_{\mu}\} & =& P_{\mu}, \label{N=4D=3twistedSUSYalgebra3}\\[2pt]
\{others\} & =& 0,\qquad \label{N=4D=3twistedSUSYalgebra4}
\end{eqnarray}
where $\epsilon_{\mu\nu\rho}$ is the three dimensional totally anti-symmetric
tensor with $\epsilon_{123}=+1$.
The Lorentz and the internal rotations 
of the supercharges are re-expressed on the
twisted basis,
\begin{eqnarray}
[J_{\mu},s] &=& +\frac{1}{2}s_{\mu}, 
\hspace{20pt}
[J_{\mu},s_{\nu}] \ =\ -\frac{i}{2}\epsilon_{\mu\nu\rho}s_{\rho}
+\frac{1}{2}\delta_{\mu\nu}s, \label{3DsJ1} \\[0pt]
[J_{\mu},\overline{s}] &=& -\frac{1}{2}\overline{s}_{\mu}, 
\hspace{20pt}
[J_{\mu},\overline{s}_{\nu}] \ =\ -\frac{i}{2}\epsilon_{\mu\nu\rho}\overline{s}_{\rho}
-\frac{1}{2}\delta_{\mu\nu}\overline{s},\label{3DsJ2} \\[0pt]
[R_{\mu},s] &=& -\frac{1}{2}s_{\mu}, 
\hspace{20pt}
[R_{\mu},s_{\nu}] \ =\ -\frac{i}{2}\epsilon_{\mu\nu\rho}s_{\rho}
-\frac{1}{2}\delta_{\mu\nu}s, \label{3DsJ3} \\[0pt]
[R_{\mu},\overline{s}] &=& +\frac{1}{2}\overline{s}_{\mu},
\hspace{20pt}
[R_{\mu},\overline{s}_{\nu}] \ =\ -\frac{i}{2}\epsilon_{\mu\nu\rho}\overline{s}_{\rho}
+\frac{1}{2}\delta_{\mu\nu}\overline{s}. \label{3DsJ4}
\end{eqnarray}
Notice that $(s,\overline{s})$ and $(s_{\mu},\overline{s}_{\mu})$ transform as
scalars and vectors
under $(SO(3)_{E}\times SO(3)_{R})_{diag}$,
respectively.
Namely, under the twisted Lorentz generator 
$J^{diag}_{\mu}=J_{\mu}+R_{\mu}$ they transform as
\begin{eqnarray}
[J^{diag}_{\mu},s] &=& [J^{diag}_{\mu},\overline{s}]\ = \ 0,\hspace{20pt} 
[J^{diag}_{\mu},s_{\nu}] \ =\ -i\epsilon_{\mu\nu\rho}s_{\rho},\hspace{20pt} 
[J^{diag}_{\mu},\overline{s}_{\nu}] \ =\ -i\epsilon_{\mu\nu\rho}\overline{s}_{\rho}.
\qquad
\end{eqnarray}

Once we have the SUSY algebra of $(s,\overline{s}_{\mu},s_{\mu},\overline{s})$, 
we can construct the corresponding
superspace by introducing the fermionic coordinates 
$\theta_{A}=(\theta, \overline{\theta}_{\mu},\theta_{\mu},\overline{\theta})$.
The $N=D=2$ and $N=D=4$ Dirac-K\"ahler twisted superspace formulations are
elaborated in \cite{KKU,KKM}.  
Here we begin by considering the following supergroup element of 
the twisted $N=4\ D=3$,
\begin{eqnarray}
G(x_{\mu},\theta,\overline{\theta}_{\mu},\theta_{\mu},\overline{\theta})
&=& e^{i(-x_{\mu}P_{\mu}+\theta s +\overline{\theta}_{\mu} \overline{s}_{\mu}
+\theta_{\mu}s_{\mu}+\overline{\theta}\overline{s})}.
\end{eqnarray}
By using the algebra 
(\ref{N=4D=3twistedSUSYalgebra1})-(\ref{N=4D=3twistedSUSYalgebra4}), 
we have,
\begin{eqnarray}
G(0,\xi,\overline{\xi}_{\mu},\xi_{\mu},\overline{\xi})
G(x_{\mu},\theta,\overline{\theta}_{\mu},\theta_{\mu},\overline{\theta})
=
G(x_{\mu}+a_{\mu},\theta+\xi,\overline{\theta}_{\mu}+\overline{\xi}_{\mu},
\theta_{\mu}+\xi_{\mu},\overline{\theta}+\overline{\xi}), \label{left_multi}
\end{eqnarray}
where the variations of the bosonic coordinates, $a_{\mu}$, are given in terms of
the combinations of the fermionic coordinates, 
\begin{eqnarray}
a_{\rho} = \frac{i}{2}\xi\overline{\theta}_{\rho}
+\frac{i}{2}\overline{\xi}\theta_{\rho}
+\frac{i}{2}\xi_{\rho}\overline{\theta}
+\frac{i}{2}\overline{\xi}_{\rho}\theta
+\frac{1}{2}\epsilon_{\mu\nu\rho}\xi_{\mu}\overline{\theta}_{\nu}
-\frac{1}{2}\epsilon_{\mu\nu\rho}\overline{\xi}_{\mu}\theta_{\nu}.
\end{eqnarray}
Expanding the r.h.s. of (\ref{left_multi}) w.r.t. 
$(\xi,\overline{\xi}_{\mu},\xi_{\mu},\overline{\xi})$,
\begin{eqnarray}
G(x_{\mu}+a_{\mu},\theta+\xi,\overline{\theta}_{\mu}+\overline{\xi}_{\mu},
\theta_{\mu}+\xi_{\mu},\overline{\theta}+\overline{\xi})
= (\xi Q + \overline{\xi}_{\mu} \overline{Q}_{\mu} + \xi_{\mu} Q_{\mu}
+\overline{\xi}\overline{Q})
G(x_{\mu},\theta,\overline{\theta}_{\mu},\theta_{\mu},\overline{\theta}), \nonumber
\\
\end{eqnarray}
one finds the superspace expressions of the $N=4\ D=3$ twisted SUSY
generators $(Q,\overline{Q}_{\mu},Q_{\mu},\overline{Q})$,
\begin{eqnarray}
Q &=& \frac{\partial}{\partial\theta} 
+ \frac{i}{2}\overline{\theta}_{\mu}\partial_{\mu},\hspace{30pt}
Q_{\mu} \ =\ \frac{\partial}{\partial\theta_{\mu}} 
+\frac{i}{2}\overline{\theta}\partial_{\mu}
+ \frac{1}{2}\epsilon_{\mu\nu\rho}\overline{\theta}_{\nu}\partial_{\rho},\\[2pt]
\overline{Q} &=& \frac{\partial}{\partial\overline{\theta}} 
+ \frac{i}{2}\theta_{\mu}\partial_{\mu}, \hspace{30pt}
\overline{Q}_{\mu} \ =\ \frac{\partial}{\partial\overline{\theta}_{\mu}} 
+ \frac{i}{2}\theta\partial_{\mu}
- \frac{1}{2}\epsilon_{\mu\nu\rho}\theta_{\nu}\partial_{\rho},
\end{eqnarray} 
which satisfy the following algebra,
\begin{eqnarray}
\{Q,\overline{Q}_{\mu}\} &=& +i\partial_{\mu}, \\[2pt]
\{Q_{\mu},\overline{Q}_{\nu}\}  &=& +\epsilon_{\mu\nu\rho}\partial_{\rho}, \\[2pt]
\{\overline{Q},Q_{\mu}\}  &=& +i\partial_{\mu},  \\[2pt]
\{others\}  &=& 0.
\end{eqnarray}
Note that the above SUSY generators are induced by the left multiplication
of the supergroup element (\ref{left_multi}).
In contrast, we also have the following $N=4\ D=3$ superderivatives 
$(D,\overline{D}_{\mu},D_{\mu},\overline{D})$
which are induced by the right multiplication of the supergroup element,
\begin{eqnarray}
G(x_{\mu},\theta,\overline{\theta}_{\mu},\theta_{\mu},\overline{\theta})
G(0,\xi,\overline{\xi}_{\mu},\xi_{\mu},\overline{\xi})
=(\xi D + \overline{\xi}_{\mu} \overline{D}_{\mu} + \xi_{\mu} D_{\mu}
+\overline{\xi}\overline{D})
G(x_{\mu},\theta,\overline{\theta}_{\mu},\theta_{\mu},\overline{\theta}),
\end{eqnarray}
\begin{eqnarray}
D &=& \frac{\partial}{\partial\theta} 
- \frac{i}{2}\overline{\theta}_{\mu}\partial_{\mu},\hspace{30pt}
D_{\mu} \ =\ \frac{\partial}{\partial\theta_{\mu}} 
-\frac{i}{2}\overline{\theta}\partial_{\mu}
- \frac{1}{2}\epsilon_{\mu\nu\rho}\overline{\theta}_{\nu}\partial_{\rho},\\[2pt]
\overline{D} &=& \frac{\partial}{\partial\overline{\theta}} 
- \frac{i}{2}\theta_{\mu}\partial_{\mu}, \hspace{30pt}
\overline{D}_{\mu} \ =\ \frac{\partial}{\partial\overline{\theta}_{\mu}} 
- \frac{i}{2}\theta\partial_{\mu}
+ \frac{1}{2}\epsilon_{\mu\nu\rho}\theta_{\nu}\partial_{\rho}.
\end{eqnarray} 
The superderivatives satisfy the following algebra,
\begin{eqnarray}
\{D,\overline{D}_{\mu}\} &=& -i\partial_{\mu}, \\[2pt]
\{D_{\mu},\overline{D}_{\nu}\}  &=& -\epsilon_{\mu\nu\rho}\partial_{\rho}, \\[2pt]
\{\overline{D},D_{\mu}\}  &=& -i\partial_{\mu}, \\[2pt]
\{others\}  &=& 0. 
\end{eqnarray}
It is important to note that 
the SUSY generators $Q_{A}=(Q,\overline{Q}_{\mu},Q_{\mu},\overline{Q})$
and the superderivatives 
$D_{A}=(D,\overline{D}_{\mu},D_{\mu},\overline{D})$
anti-commute each other,
\begin{eqnarray}
\{Q_{A},D_{B}\} &=& 0.
\end{eqnarray}

Having these superspace operators in hand,
we then proceed to formulate the $N=4\ D=3$ twisted SYM
by means of the superfield method.
One of the most systematic treatments to construct the 
supersymmetric gauge theories is the so-called super-connection method 
which was introduced in \cite{WZ_GSW_S}.
It was also applied to the investigation of topological quantum field theory
\cite{Alvarez_Labastida}.
The detailed analysis of 
super-connection formulations for twisted $N=D=2$ and $N=2\ D=4$
from the Dirac-K\"ahler point of view are given in \cite{KKM,KM}. 

We first introduce the $N=4\ D=3$ fermionic gauge covariant derivatives
$\D_{A}=(\D,\overline{\D}_{\mu},\D_{\mu},\overline{\D})$,
\begin{eqnarray}
\D &=& D -i\Gamma(x,\theta,\overline{\theta}_{\mu},\theta_{\mu},\overline{\theta}),
\hspace{20pt}
\D_{\mu} \ =\ D_{\mu} -i\Gamma_{\mu}
(x,\theta,\overline{\theta}_{\mu},\theta_{\mu},\overline{\theta}), \\[2pt]
\overline{\D} &=& \overline{D} 
-i\overline{\Gamma}(x,\theta,\overline{\theta}_{\mu},\theta_{\mu},\overline{\theta}),
\hspace{20pt}
\overline{\D}_{\mu} \ =\ \overline{D}_{\mu} -i\overline{\Gamma}_{\mu}
(x,\theta,\overline{\theta}_{\mu},\theta_{\mu},\overline{\theta}), 
\end{eqnarray}
where $\Gamma_{A}=(\Gamma,\overline{\Gamma}_{\mu},\Gamma_{\mu},\overline{\Gamma})$
are denoting the superconnections
associated with the superderivatives.
All of the $\Gamma_{A}$'s
are the functions of 
$(x,\theta,\overline{\theta}_{\mu},\theta_{\mu},\overline{\theta})$
and are belonging to the adjoint representation of the gauge group.
The fermionic gauge covariant derivatives $\D_{A}$ are transforming 
under the supergauge transformations as follows,
\begin{eqnarray}
\D_{A} \rightarrow \D_{A}'=e^{-i\Omega}\D_{A}e^{+i\Omega}, \label{gauge_trans_nabla}
\end{eqnarray}
where $\Omega=\Omega(x,\theta,\overline{\theta}_{\mu},\theta_{\mu},\overline{\theta})$
denotes the generic hermitian superfield as we will see later on. 
Since the above super-connections $\Gamma_{A}$ contain 
a large number of component fields,
\begin{eqnarray}
\Gamma_{A} &=& \psi_{A} + \theta_{B}\psi_{BA} + \theta_{C}\theta_{B}\psi_{CBA} + \cdots ,
\end{eqnarray}
the resulting multiplet might become highly reducible in general 
even after taking the Wess-Zumino gauge.
The central issue of formulating the extended supersymmetric gauge theories
is thus how to reduce the number of component fields 
and how to obtain the irreducible SUSY multiplet in a gauge covariant manner.

One of the possible ways to obtain such an irreducible $N=4\ D=3$ twisted SYM
multiplet is to impose the following constraints on the 
fermionic gauge covariant derivatives,
\begin{eqnarray}
\{\D,\overline{\D}_{\mu}\} &=& -i(\D_{\underline{\mu}}-i\Phi^{(\mu)}),
\label{N=4D=3const1} \\[2pt]
\{\D_{\mu},\overline{\D}_{\nu}\} &=& 
-\epsilon_{\mu\nu\rho}(\D_{\underline{\rho}}+i\Phi^{(\rho)}), 
\label{N=4D=3const2}\\[2pt]
\{\overline{\D},{\D}_{\mu}\} &=& -i(\D_{\underline{\mu}}-i\Phi^{(\mu)}), 
\label{N=4D=3const3}\\[2pt]
\{others\} &=& 0.\label{N=4D=3const4}
\end{eqnarray}
The $\D_{\underline{\mu}}$ and $\Phi^{(\mu)}$ are the superfields
whose lowest components are representing the gauge covariant derivatives
and the scalar fields, respectively,
\begin{eqnarray}
\D_{\underline{\mu}} &=& \partial_{\mu}-iA_{\mu} + \cdots,\\[2pt]
\Phi^{(\mu)} &=& \phi^{(\mu)} + \cdots,
\end{eqnarray}
here and in the following the dots $\cdots$ are representing 
the possible higher order terms w.r.t.
the fermionic coordinates 
$\theta_{A}=(\theta,\overline{\theta}_{\mu},\theta_{\mu},\overline{\theta})$.
All of the components in the superfields $\D_{\underline{\mu}}$ and $\Phi^{(\mu)}$,
including $A_{\mu}$ and $\phi^{(\mu)}$,
can be essentially expressed in terms of the component fields 
embedded in the superconnections
$\Gamma_{A}=(\Gamma,\overline{\Gamma}_{\mu},\Gamma_{\mu},\overline{\Gamma})$.

There are several remarks in order.
First, since in the constraints (\ref{N=4D=3const1})-(\ref{N=4D=3const4})
we introduced the scalar fields $\phi^{(\mu)}$
on the same footing as the gauge fields $A_{\mu}$, 
one may wonder how these fields are transformed under the 
Lorentz $SO(3)_{E}$ and the internal $SO(3)_{R}$ rotations.  
Reminding that the fermionic covariant derivatives 
$\D_{A}=(\D,\overline{\D}_{\mu},\D_{\mu},\overline{\D})$
are transforming just like as $s_{A}=(s,\overline{s}_{\mu},s_{\mu},\overline{s})$
in (\ref{3DsJ1})-(\ref{3DsJ4}), respectively, 
one finds from (\ref{N=4D=3const1})-(\ref{N=4D=3const4}),
\begin{eqnarray}
[J_{\mu},\D_{\underline{\nu}} \pm i\Phi^{(\nu)}]
&=& -\frac{i}{2}\epsilon_{\mu\nu\rho}(\D_{\underline{\rho}}+i\Phi^{(\rho)})
-\frac{i}{2}\epsilon_{\mu\nu\rho}(\D_{\underline{\rho}}-i\Phi^{(\rho)}) \\[2pt]
&=& -i\epsilon_{\mu\nu\rho} \D_{\underline{\rho}}, 
\end{eqnarray}
from which it obeys,
\begin{eqnarray}
[J_{\mu},\D_{\underline{\nu}}] &=& -i\epsilon_{\mu\nu\rho}\D_{\underline{\rho}},
\hspace{30pt}
[J_{\mu},\Phi^{(\nu)}] \ =\ 0. 
\end{eqnarray}
Namely, the gauge fields $A_{\mu}$ are actually transforming as a
$SO(3)_{E}$ vector while the scalar fields $\phi^{(\mu)}$
are transforming as $SO(3)_{E}$ scalars.
In contrast, one could also obtain
\begin{eqnarray}
[R_{\mu},\D_{\underline{\nu}}] &=& 0,
\hspace{30pt}
[R_{\mu},\Phi^{(\nu)}] \ =\ -i\epsilon_{\mu\nu\rho}\Phi^{(\rho)}, 
\end{eqnarray}
which implies that the $A_{\mu}$ and $\phi^{(\mu)}$ 
are transforming as $SO(3)_{R}$ scalars and a vector,
respectively.
One can thus see that even though the $\phi^{(\mu)}$ are
introduced on the same footing as the gauge fields $A_{\mu}$, 
they are appropriately transforming as scalars of the original Lorentz rotations.
Obviously, the sign difference in front of the $\Phi^{(\mu)}$
in (\ref{N=4D=3const1})-(\ref{N=4D=3const3}) is responsible for these
transformation properties. 
Furthermore, one should notice that, after the twisting, 
both of $A_{\mu}$ and $\phi^{(\mu)}$  
transform as vectors under the twisted rotational group
$(SO(3)_{E}\times SO(3)_{R})_{diag}$,
\begin{eqnarray}
[J^{diag}_{\mu},\D_{\underline{\nu}}] &=& -i\epsilon_{\mu\nu\rho}\D_{\underline{\rho}},
\hspace{30pt}
[J^{diag}_{\mu},\Phi^{(\nu)}] \ =\ -i\epsilon_{\mu\nu\rho}\Phi^{(\rho)}. 
\end{eqnarray}
The combinations $\D_{\underline{\mu}}\mp i\Phi^{(\mu)}$
appeared in the r.h.s. of (\ref{N=4D=3const1})-(\ref{N=4D=3const3})
are thus the covariant expressions w.r.t. the twisted rotational group
$(SO(3)_{E}\times SO(3)_{R})_{diag}$.

The second remark is regarding the hermiticity of the constraints
(\ref{N=4D=3const1})-(\ref{N=4D=3const4}).
We impose the following hermitian conjugation properties on the
fermionic covariant derivatives,
\begin{eqnarray}
\D^{\dagger} &=& \overline{\D}, \hspace{30pt}
\D_{\mu}^{\dagger} \ =\ \overline{\D}_{\mu},
\end{eqnarray}
which are consistent with the complex conjugation nature of the 
supercharges $Q_{\alpha i}$ and $\overline{Q}_{i \alpha}$ in
(\ref{N=4D=3SUSYalgebra}),
$Q_{\alpha i}^{*}=\overline{Q}_{i \alpha}$.
One could easily notice that, in order to be compatible 
with the constraints (\ref{N=4D=3const1})-(\ref{N=4D=3const3}),
the supergauge transformation $\Omega$ in (\ref{gauge_trans_nabla})
should be hermitian.
Correspondingly, one can take the $A_{\mu}$ and the scalars 
$\phi^{(\mu)}$ as the hermitian fields which transform under the gauge 
transformation as,
\begin{eqnarray}
\partial_{\mu}-iA_{\mu} &\rightarrow& 
e^{-i\omega}(\partial_{\mu}-iA_{\mu})e^{+i\omega},\hspace{30pt}
\phi^{(\mu)} \ \rightarrow\  e^{-i\omega}\phi^{(\mu)}e^{+i\omega},
\end{eqnarray}
where $\omega$ denotes the $\theta_{A}$ independent first component of 
the supergauge transformation $\Omega$
satisfying $\omega^{\dagger}=\omega$.
 
The third remark is that
the constraints (\ref{N=4D=3const1})-(\ref{N=4D=3const4})
corresponds to the naive continuum limit of the $N=4\ D=3$ lattice SYM
constraints recently proposed in \cite{DKKN3},
where the gauge fields $A_{\mu}$ are exponentiated
together with the scalar fields $\phi^{(\mu)}$
such that they could represent the bosonic gauge link variables
either of the forward or the backward type,
\begin{eqnarray}
\partial_{\mu}-i(A_{\mu}\pm \phi^{(\mu)})
\rightarrow \mp (e^{\pm i(A_{\mu}\pm \phi^{(\mu)})})_{x\pm n_{\mu},x},
\label{bosonic_link_variables}
\end{eqnarray} 
where the subscripts indicate
that they are located on links from $x$ to $x + n_{\mu}$ (forward) 
and $x$ to $x-n_{\mu}$ (backward), respectively, for a generic site $x$.
As is also stressed in \cite{DKKN1,DKKN2,DKKN3},
the twisting is playing a fundamental role in realizing the 
supersymmetry on the lattice and
it can be traced back to the intrinsic relation between
the twisted fermions and the Dirac-K\"ahler fermions
\cite{KT,KKU,KKM}.
Here we find the importance of 
the twisting in the bosonic sector as well.
Namely, the exponential forms in (\ref{bosonic_link_variables})
can transform covariantly only under the twisted rotational group
$(SO(3)_{E}\times SO(3)_{R})_{diag}$
and not under the $SO(3)_{E}$ and $SO(3)_{R}$ independently.
We will come back to this point once again 
in Sec. \ref{lattice_realization_other_types}.

Once we impose the constraints (\ref{N=4D=3const1})-(\ref{N=4D=3const4}),
the whole information of the resulting $N=4\ D=3$ twisted SYM multiplet
can be obtained by analyzing the Jacobi identities together with the
constraints (\ref{N=4D=3const1})-(\ref{N=4D=3const4}).
For the notational simplicity, we re-write the constraints  
(\ref{N=4D=3const1})-(\ref{N=4D=3const4}) as
\begin{eqnarray}
\{\D,\overline{\D}_{\mu}\} &=& -i\D_{+\mu},
\label{N=4D=3const1r} \\[2pt]
\{\D_{\mu},\overline{\D}_{\nu}\} &=& 
-\epsilon_{\mu\nu\rho}\D_{-\rho}, 
\label{N=4D=3const2r}\\[2pt]
\{\overline{\D},{\D}_{\mu}\} &=& -i\D_{+\mu}, 
\label{N=4D=3const3r}\\[2pt]
\{others\} &=& 0.\label{N=4D=3const4r}
\end{eqnarray}
The symbols $\D_{\pm\mu}$ are defined by
\begin{eqnarray}
\D_{\pm\mu} &=& \D_{\underline{\mu}}\mp i\Phi^{(\mu)} \\[2pt]
&=& \partial_{\mu}-i(A_{\mu}\pm \phi^{(\mu)}) + \cdots, \\[2pt]
&=& \Dc_{\pm\mu} +\cdots,
\end{eqnarray}
where we denote the $\theta_{A}$ independent part of 
$\D_{\pm\mu}$
as $\Dc_{\pm\mu}=\partial_{\mu}-i(A_{\mu}\pm \phi^{(\mu)})$.
Since the $N=4\ D=3$ SYM constraints (\ref{N=4D=3const1r})-(\ref{N=4D=3const4r})
are formally similar to the lattice SYM constraints in \cite{DKKN3},
the Jacobi identity analysis also goes parallel to the lattice analysis
\footnote{Since in the lattice formulations of SYM \cite{DKKN2,DKKN3}
all the operators are generically defined on links,
all the (anti-)commutators are replaced by the ``link" (anti-)commutators.
See also the Sec. \ref{lattice_realization_other_types} of this paper.}.
The Jacobi identities of three fermionic covariant derivatives give,
\begin{eqnarray}
[\D_{\mu},\D_{+\nu}] + [\D_{\nu},\D_{+\mu}] &=& 0,  \hspace{57pt}
[\overline{\D}_{\mu},\D_{+\nu}]
+ [\overline{\D}_{\nu},\D_{+\mu}] \ = \ 0, \qquad \label{Jacobi1}\\[2pt]
[\D_{\mu},\D_{+\nu}]
 -i\epsilon_{\mu\nu\rho}[\D,\D_{-\rho}] &=& 0, \hspace{40pt} 
[\overline{\D}_{\mu},\D_{+\nu}]
 +i\epsilon_{\mu\nu\rho}[\overline{\D},\D_{-\rho}] \ =\ 0, \label{Jacobi2}\\[2pt]
\epsilon_{\mu\nu\lambda}[\D_{\rho},\D_{-\lambda}]
+\epsilon_{\rho\nu\lambda}[\D_{\mu},\D_{-\lambda}] &=& 0, \hspace{19pt} 
\epsilon_{\mu\nu\lambda}[\overline{\D}_{\rho},\D_{-\lambda}]
+\epsilon_{\rho\nu\lambda}[\overline{\D}_{\mu},\D_{-\lambda}] \ =\ 0,
\label{Jacobi3} \\[2pt]
[\D,\D_{+\mu}] \ =\   
[\overline{\D},\D_{+\mu}] &=& 0, \label{Jacobi4}   
\end{eqnarray} 
from which we can define the following non-vanishing fermionic 
superfields $(\Upsilon,\overline{\Lambda}_{\mu},\Lambda_{\mu},\overline{\Upsilon})$
\footnote{For the later convenience, we took the sign conventions of 
$(\rho,\orho)$ oppositely from the
ones given in \cite{DKKN3}.},
\begin{eqnarray}
[\D,\D_{-\rho}] &\equiv& +(\overline{\Lambda}_{\rho}), 
\hspace{49pt}
[\overline{\D},\D_{-\rho}] \ \equiv\  +(\Lambda_{\rho}), \label{fermion1} \\[2pt]
[\D_{\mu},\D_{+\nu}] &=&
+i\epsilon_{\mu\nu\rho}(\overline{\Lambda}_{\rho}), 
\hspace{20pt}
[\overline{\D}_{\mu},\D_{+\nu}] \ =\
-i\epsilon_{\mu\nu\rho}(\Lambda_{\rho}), \label{fermion2} \\[2pt]
[\D_{\mu},\D_{-\nu}]
&\equiv& -\delta_{\mu\nu}(\overline{\Upsilon}), 
\hspace{32pt}
[\overline{\D}_{\mu},\D_{-\nu}] 
\ \equiv\  -\delta_{\mu\nu}(\Upsilon). \label{fermion3}
\end{eqnarray}
We denote the lowest components of the fermionic superfields
$(\Upsilon,\overline{\Lambda}_{\mu},\Lambda_{\mu},\overline{\Upsilon})$ as
$(\rho,\overline{\lambda}_{\mu},\lambda_{\mu},\overline{\rho})$,
\begin{eqnarray}
\Upsilon &=& \rho +\cdots, \hspace{20pt}
\overline{\Upsilon} \ =\ \overline{\rho} +\cdots, \hspace{20pt}
\Lambda_{\mu} \ =\ \lambda_{\mu} + \cdots, \hspace{20pt}
\overline{\Lambda}_{\mu} \ =\ \overline{\lambda}_{\mu} + \cdots. 
\end{eqnarray}
The $(\rho,\overline{\lambda}_{\mu},\lambda_{\mu},\overline{\rho})$
are representing the $N=4\ D=3$ twisted fermions in the SYM multiplet.
The vanishing conditions resulting from the
relations (\ref{Jacobi1})-(\ref{Jacobi4})
also give rise to the covariant ``chiral" or ``anti-chiral" conditions
for $\D_{\pm\mu}$, for example,
\begin{eqnarray}
[\D,\D_{+3}] \ =\ [\oD_{3},\D_{+3}] \ =\ [\D_{3},\D_{+3}] \ =\
[\oD,\D_{+3}] &=& 0,\label{chiral_cond1} \\[2pt]
[\D_{1},\D_{-3}] \ =\ [\D_{2},\D_{-3}] \ =\ 
[\oD_{1},\D_{-3}] \ =\ [\oD_{2},\D_{-3}] &=& 0. \label{chiral_cond2}
\end{eqnarray} 
One also has the similar conditions for $\D_{\pm 1}$ and $\D_{\pm 2}$.
All the commutators of $\D_{A}=(\D,\oD_{\mu},\D_{\mu},\oD)$ and 
$\D_{\pm\mu}$ are summarized in Table \ref{commutators}.
As we will see, these conditions are playing important roles when constructing 
the twisted SUSY invariant action.

\begin{table}
\begin{center}
\renewcommand{\arraystretch}{1.25}
\renewcommand{\tabcolsep}{5pt}
\begin{tabular}{|c|c|c|c|c|c|c|c|c|}
\hline
 & $\D$ & $\oD_{1}$ & $\oD_{2}$ & $\oD_{3}$ &
$\D_{1}$ & $\D_{2}$ & $\D_{3}$ & $\oD$ \\  \hline 
$\D_{+1}$ & $0$ & $0$ & $+i\Lambda_{3}$ & $-i\Lambda_{2}$
& $0$ & $-i\oLambda_{3}$ & $+i\oLambda_{2}$ & $0$ \\ 
$\D_{-1}$ & $+\oLambda_{1}$ & $-\Upsilon$ & $0$ & $0$
& $-\oUpsilon$ & $0$ & $0$ & $+\Lambda_{1}$ \\ \hline
$\D_{+2}$ & $0$ & $-i\Lambda_{3}$ & $0$ & $+i\Lambda_{1}$
& $+i\oLambda_{3}$ & $0$ & $-i\oLambda_{1}$ & $0$ \\ 
$\D_{-2}$ & $+\oLambda_{2}$ & $0$ & $-\Upsilon$ & $0$
& $0$ & $-\oUpsilon$ & $0$ & $+\Lambda_{2}$ \\ \hline
$\D_{+3}$ & $0$ & $+i\Lambda_{2}$ & $-i\Lambda_{1}$ & $0$
& $-i\oLambda_{2}$ & $+i\oLambda_{1}$ & $0$ & $0$ \\ 
$\D_{-3}$ & $+\oLambda_{3}$ & $0$ & $0$ & $-\Upsilon$
& $0$ & $0$ & $-\oUpsilon$ & $+\oLambda_{3}$ \\ \hline  
\end{tabular}
\caption{All components of the commutators 
 $[\D_{A},\D_{\pm\mu}]$ with
 $\D_{A}=(\D,\oD_{\mu},\D_{\mu},\oD)$ }
\label{commutators}
\end{center}
\end{table}

By taking the anti-commutators of 
$\D_{A}$'s with the relations (\ref{fermion1})-(\ref{fermion3}),
we have,
\begin{eqnarray}
\{\D,\Lambda_{\mu}\}-\frac{1}{2}\epsilon_{\mu\rho\sigma}[\D_{+\rho},\D_{+\sigma}]
\ =\ 0,  \hspace{20pt}
\{\overline{\D},\overline{\Lambda}_{\mu}\}
+\frac{1}{2}\epsilon_{\mu\rho\sigma}[\D_{+\rho},\D_{+\sigma}]
& =& 0, \\[2pt]
\{\overline{\D}_{\mu},\Lambda_{\nu}\}
-\delta_{\mu\nu}\{\overline{\D},\Upsilon\} 
\ =\ 0, \hspace{46pt}
\{\D_{\mu},\overline{\Lambda}_{\nu}\}
-\delta_{\mu\nu}\{\D,\overline{\Upsilon}\} 
& =& 0, \\[4pt]
\epsilon_{\lambda\nu\rho}\{\overline{\D}_{\mu},\overline{\Lambda}_{\rho}\}
+i\epsilon_{\lambda\mu\rho}[\D_{+\nu},\D_{-\rho}]
-\epsilon_{\mu\nu\rho}\{\D_{\lambda},\Lambda_{\rho}\} &=& 0, \\[4pt]
\delta_{\mu\nu}\{\overline{\D},\overline{\Upsilon}\}
+i[\D_{-\nu},\D_{+\mu}]
-\{\D_{\mu},\Lambda_{\nu}\} &=& 0, \\[4pt]
\delta_{\mu\nu}\{\D,\Upsilon\}
+i[\D_{-\nu},\D_{+\mu}]
-\{\overline{\D}_{\mu},\overline{\Lambda}_{\nu}\} &=& 0, \\[4pt]
\delta_{\lambda\nu}\{\overline{\D}_{\mu},\overline{\Upsilon}\}
+\epsilon_{\lambda\mu\rho}[\D_{-\nu},\D_{-\rho}]
+\delta_{\mu\nu}\{\D_{\lambda},\Upsilon\} &=& 0, \\[4pt]
\{\overline{\D},\Lambda_{\mu}\} \ =\ 
\{\D,\overline{\Lambda}_{\mu}\} \ =\ 
\{\overline{\D}_{\mu},\Upsilon\} \ =\
\{\D_{\mu},\overline{\Upsilon}\} & =& 0,
\end{eqnarray} 
which can be solved w.r.t. the anti-commutators
of the fermionic derivatives $\D_{A}$ and 
the fermionic superfields $(\Lambda_{\mu},\oLambda_{\mu})$,
\begin{eqnarray}
\{\D,\Lambda_{\mu}\} &=& +\frac{1}{2}\epsilon_{\mu\rho\sigma}
[\D_{+\rho},\D_{+\sigma}], \hspace{30pt} 
\{\oD,\oLambda_{\mu}\} \ =\ -\frac{1}{2}\epsilon_{\mu\rho\sigma}
[\D_{+\rho},\D_{+\sigma}], \\[2pt]
\{\D_{\mu},\oLambda_{\nu}\} &=& -\delta_{\mu\nu}\cG, \hspace{84pt} 
\{\oD_{\mu},\Lambda_{\nu}\} \ =\ -\delta_{\mu\nu}\ocG, \\[2pt]
\{\D_{\mu},\Lambda_{\nu}\} &=& -i[\D_{+\mu},\D_{-\nu}]
-\delta_{\mu\nu}(\cK-\frac{i}{2}[\D_{+\rho},\D_{-\rho}]), \qquad \\[0pt]
\{\oD_{\mu},\oLambda_{\nu}\}
 &=& -i[\D_{+\mu},\D_{-\nu}]
+\delta_{\mu\nu}(\cK+\frac{i}{2}[\D_{+\rho},\D_{-\rho}]), \\[2pt]
\{\D,\oLambda_{\mu}\} &=& \{\oD,\Lambda_{\mu}\} \ =\ 0,
\end{eqnarray}
and of the fermionic derivatives $\D_{A}$ and the superfields $(\Upsilon,\oUpsilon)$,
\begin{eqnarray}
\{\D_{\mu},\Upsilon\} &=& -\frac{1}{2}\epsilon_{\mu\rho\sigma}
[\D_{-\rho},\D_{-\sigma}], \hspace{30pt} 
\{\oD_{\mu},\oUpsilon\} \ =\ +\frac{1}{2}\epsilon_{\mu\rho\sigma}
[\D_{-\rho},\D_{-\sigma}], \\[0pt]
\{\D,\Upsilon\} &=& +\cK+\frac{i}{2}[\D_{+\rho},\D_{-\rho}], \hspace{31pt} 
\{\oD,\oUpsilon\} \ =\ -\cK+\frac{i}{2}[\D_{+\rho},\D_{-\rho}], \\[0pt]
\{\D,\oUpsilon\} &=& -\cG, \hspace{111pt} 
\{\oD,\Upsilon\} \ =\ -\ocG, \\[5pt]
\{\D_{\mu},\oUpsilon\} &=& 
\{\oD_{\mu},\Upsilon\} \ =\ 0, 
\end{eqnarray}
where we introduced the auxiliary superfield
$\cG$, $\ocG$, $\cK$ whose first components are representing
the bosonic auxiliary fields $(G,\oG,K)$ in the $N=4\ D=3$ twisted SYM multiplet,
\begin{eqnarray}
\cG &=& G +\cdots,
\hspace{30pt}
\ocG \ =\  \oG +\cdots,
\hspace{30pt}
\cK \ =\ K +\cdots.
\end{eqnarray}
One can show that all the other higher Jacobi identities can be 
expressed in terms of the fermionic covariant derivatives 
$(\D,\oD_{\mu},\D_{\mu},\oD)$,
the gauge covariant derivative superfields $\D_{\pm\mu}$,
the non-vanishing fermionic superfields 
$(\Upsilon,\oLambda_{\mu},\Lambda_{\mu},\oUpsilon)$ 
and the auxiliary superfields $(\cG,\ocG,\cK)$.
As we will see, the lowest components 
of the superfields
$(\D_{\pm\mu},\Upsilon,\oLambda_{\mu},\Lambda_{\mu},\oUpsilon,
\cG,\ocG,\cK)$
are representing the off-shell multiplet of $N=4\ D=3$ twisted SYM.

The $SU(2)_{E}\times SU(2)_{R}$ rotational properties of the
component fields can also be read off from the above Jacobi identities,
\begin{eqnarray}
[J_{\mu},\rho] &=& +\frac{1}{2}\lambda_{\mu}, \hspace{30pt}
[J_{\mu},\lambda_{\nu}]\ =\ -\frac{i}{2}\epsilon_{\mu\nu\rho}\lambda_{\rho}
+\frac{1}{2}\delta_{\mu\nu}\rho, \\[2pt] 
[J_{\mu},\orho] &=& -\frac{1}{2}\lambda_{\mu}, \hspace{30pt}
[J_{\mu},\olambda_{\nu}]\ =\ -\frac{i}{2}\epsilon_{\mu\nu\rho}\olambda_{\rho}
-\frac{1}{2}\delta_{\mu\nu}\orho, \\[2pt] 
[R_{\mu},\rho] &=& -\frac{1}{2}\lambda_{\mu}, \hspace{30pt}
[R_{\mu},\lambda_{\nu}]\ =\ -\frac{i}{2}\epsilon_{\mu\nu\rho}\lambda_{\rho}
-\frac{1}{2}\delta_{\mu\nu}\rho, \\[2pt] 
[R_{\mu},\orho] &=& +\frac{1}{2}\lambda_{\mu}, \hspace{30pt}
[R_{\mu},\olambda_{\nu}]\ =\ -\frac{i}{2}\epsilon_{\mu\nu\rho}\olambda_{\rho}
+\frac{1}{2}\delta_{\mu\nu}\orho, \\[4pt] 
[J_{\mu},G] &=& [J_{\mu},\oG] \ =\ [J_{\mu},K] \ =\ 
[R_{\mu},G] \ =\ [R_{\mu},\oG] \ =\ [R_{\mu},K] \ =\ 0.
\end{eqnarray}
One sees that, after the twisting, 
the fermions $(\rho,\olambda_{\mu},\lambda_{\mu},\orho)$
are transforming as (scalar, vector, vector, scalar) while
all of the auxiliary fields $(G,\oG,K)$ remain as scalars
under $J^{diag}_{\mu}=J_{\mu}+R_{\mu}$.

The SUSY transformations of the component fields can be determined from the
above Jacobi identities via
\begin{eqnarray}
s_{A}\varphi &=& \{\D_{A},\Psi]|_{\theta's=0},
\end{eqnarray}
where the $\varphi$ denotes any of the component field
$(\Dc_{\pm\mu},\rho,\olambda_{\mu},\lambda_{\mu},\orho,G,\oG,K)$
in the SYM multiplet
while the $\Psi$ denotes the corresponding superfields 
$(\D_{\pm\mu},\Upsilon,\oLambda_{\mu},\Lambda_{\mu},\oUpsilon,
\mathcal{G},\overline{\mathcal{G}},\mathcal{K})$, respectively. 
The symbol $|_{\theta's=0}$ means that the 
$\theta_{A}=(\theta,\overline{\theta}_{\mu},\theta_{\mu},\overline{\theta})$
are all taken to be zero.
All the $N=4\ D=3$ twisted SUSY transformation laws for the component fields
are listed in Table \ref{N=4D=3SYMtrans}.
As a natural consequence of the constraints 
(\ref{N=4D=3const1})-(\ref{N=4D=3const4}) or 
(\ref{N=4D=3const1r})-(\ref{N=4D=3const4r}),
the resulting $N=4\ D=3$ twisted SUSY algebra for the
component fields closes off-shell modulo gauge transformations,
\begin{eqnarray}
\{s,\os_{\mu}\}\varphi 
&=& -i[\Dc_{+\mu},\varphi], \label{N=4D=3SYMalgebra1}\\[2pt]
\{s_{\mu},\os_{\nu}\}\varphi
&=& -\epsilon_{\mu\nu\rho}[\Dc_{-\rho},\varphi],
\label{N=4D=3SYMalgebra2}\\[2pt]
\{\os, s_{\mu}\}\varphi
&=& -i[\Dc_{+\mu},\varphi],\label{N=4D=3SYMalgebra3}\\[2pt]
\{others\}\varphi &=& 0, \label{N=4D=3SYMalgebra4}
\end{eqnarray}
where the $\varphi$ again denotes any component of the SYM multiplet
$(\Dc_{\pm\mu},\rho,\olambda_{\mu},\lambda_{\mu},\orho,G,\oG,K)$.
As described in \cite{KKU,KKM,WB}, once all the SUSY transformation laws 
of the component fields are obtained, the corresponding superfield expressions 
can be given by operating $e^{\delta_{\theta}}$ on the lowest components,
where $\delta_{\theta}=\theta s + \overline{\theta}_{\mu}\os_{\mu}
+\theta_{\mu}s_{\mu}+\overline{\theta}\os$ 
for the twisted $N=4\ D=3$,
\begin{eqnarray}
(\D_{\pm\mu},\Upsilon,\oLambda_{\mu},\Lambda_{\mu},\oUpsilon,
\mathcal{G},\mathcal{\oG},\mathcal{K}) &=&
e^{\delta_{\theta}}(\Dc_{\pm\mu},\rho,\olambda_{\mu},\lambda_{\mu},\orho,G,\oG,K).
\end{eqnarray}

\begin{table}
\begin{center}
\renewcommand{\arraystretch}{1.25}
\renewcommand{\tabcolsep}{3pt}
\begin{tabular}{|c|c|c|c|c|}
\hline
& $s$ & $\os_{\mu}$ & $s_{\mu}$ & $\os$ \\ \hline
$\Dc_{+\nu}$ & $0$ & $-i\epsilon_{\mu\nu\rho}\lambda_{\rho}$
& $+i\epsilon_{\mu\nu\rho}\olambda_{\rho}$ & $0$ \\ 
$\Dc_{-\nu}$ & $+\olambda_{\nu}$ & $-\delta_{\mu\nu}\rho$ 
& $-\delta_{\mu\nu}\orho$ & $+\lambda_{\nu}$ \\ \hline
$\rho$ & $+K+\frac{i}{2}[\Dc_{+\rho},\Dc_{-\rho}]$ & $0$ 
& $-\frac{1}{2}\epsilon_{\mu\rho\sigma}[\Dc_{-\rho},\Dc_{-\sigma}]$ & $-\oG$ \\
$\olambda_{\nu}$ & $0$ 
& $-i[\Dc_{+\mu},\Dc_{-\nu}]$ 
& $-\delta_{\mu\nu}G$ 
& $-\frac{1}{2}\epsilon_{\nu\rho\sigma}[\Dc_{+\rho},\Dc_{+\sigma}]$
\\ 
& & $+\delta_{\mu\nu}(K+\frac{i}{2}[\Dc_{+\rho},\Dc_{-\rho}])$ & & \\
$\lambda_{\nu}$ 
& $+\frac{1}{2}\epsilon_{\nu\rho\sigma}[\Dc_{+\rho},\Dc_{+\sigma}]$ 
& $-\delta_{\mu\nu}\oG$ 
& $-i[\Dc_{+\mu},\Dc_{-\nu}]$ 
& $0$  
\\ 
& & & $-\delta_{\mu\nu}(K-\frac{i}{2}[\Dc_{+\rho},\Dc_{-\rho}])$ & \\
$\orho$ & $-G$ 
& $+\frac{1}{2}\epsilon_{\mu\rho\sigma}[\Dc_{-\rho},\Dc_{-\sigma}]$ 
& $0$ & $-K+\frac{i}{2}[\Dc_{+\rho},\Dc_{-\rho}]$ \\ \hline
$G$ & $0$ 
& $+\epsilon_{\mu\rho\sigma}[\Dc_{-\rho},\olambda_{\sigma}]$ 
& $0$
& $+i[\Dc_{+\rho},\olambda_{\rho}]$  
\\ 
& & $+i[\Dc_{+\mu},\orho]$ & &  \\
$\oG$ & $+i[\Dc_{+\rho},\lambda_{\rho}]$ & $0$ 
& $-\epsilon_{\mu\rho\sigma}[\Dc_{-\rho},\lambda_{\sigma}]$ & $0$ 
\\ 
& & & $+i[\Dc_{+\mu},\rho]$ &  \\
$K$ & $-\frac{i}{2}[\Dc_{+\rho},\olambda_{\rho}]$ 
& $+\frac{1}{2}\epsilon_{\mu\rho\sigma}[\Dc_{-\rho},\lambda_{\sigma}]$
& $+\frac{1}{2}\epsilon_{\mu\rho\sigma}[\Dc_{-\rho},\olambda_{\sigma}]$
& $+\frac{i}{2}[\Dc_{+\rho},\lambda_{\rho}]$ 
\\ 
& & $-\frac{i}{2}[\Dc_{+\mu},\rho]$ & $+\frac{i}{2}[\Dc_{+\mu},\orho]$ & \\ 
\hline 
\end{tabular}
\caption{SUSY trans. laws for twisted $N=4\ D=3$ SYM multiplet
$(\Dc_{\pm\mu},\rho,\olambda_{\mu},\lambda_{\mu},\orho,G,\oG,K)$}
\label{N=4D=3SYMtrans}
\end{center}
\end{table}

The $N=4\ D=3$ twisted SYM action can be manifestly constructed 
with the help of the ``chiral" and ``anti-chiral" superfields $\D_{\pm\mu}$.
For example, if we focus on the
$\D_{\pm 3}$ which are 
subject to (\ref{chiral_cond1})-(\ref{chiral_cond2}),
the invariant action can be given by either of the following expressions,
\begin{eqnarray}
\int d^{3}x \int d\overline{\theta}_{1}d\overline{\theta}_{2}
d\theta_{1} d\theta_{2} \ {\mathrm{tr}} \D_{+3}\D_{+3}, \hspace{20pt}
\int d^{3}x \int d\overline{\theta} d\theta
d\overline{\theta}_{3} d\theta_{3} 
\ {\mathrm{tr}} \D_{-3}\D_{-3}. 
\end{eqnarray}
In terms of the lowest component fields $\Dc_{\pm 3}$, the above expressions are
essentially equivalent to the successive operations of the supercharges
on the lowest components $\Dc_{\pm 3}$,
\begin{eqnarray}
\int d^{3}x \int \overline{s}_{1}\overline{s}_{2}
s_{1} s_{2} \ {\mathrm{tr}} \Dc_{+3}\Dc_{+3}, \hspace{20pt}
\int d^{3}x \int \overline{s} s
\overline{s}_{3} s_{3} 
\ {\mathrm{tr}} \Dc_{-3}\Dc_{-3}, 
\end{eqnarray}
respectively.
By consulting the SUSY transformation laws summarized in Table \ref{N=4D=3SYMtrans},
one can show that these two combinations are equivalent and 
give rise to the following SYM action,
\begin{eqnarray}
S^{N=4\ D=3}_{TSYM} &=& \int d^{3}x\ \frac{1}{2}\os_{1}\os_{2}s_{1}s_{2}
\ \mathrm{tr}\ \Dc_{+3}\ \Dc_{+3} 
\ =\ \int d^{3}x\ \frac{1}{2}s \os \os_{3} s_{3}
\ \mathrm{tr}\ \Dc_{-3}\ \Dc_{-3} \label{SYMaction_1stline} \\ 
&=& \int d^{3}x \ \mathrm{tr}\biggl[\
\frac{1}{4}[\Dc_{+\mu},\Dc_{-\mu}][\Dc_{+\nu},\Dc_{-\nu}] 
-\frac{1}{2}
[\Dc_{+\mu},\Dc_{+\nu}]
[\Dc_{-\mu},\Dc_{-\nu}] 
+K^{2}  +G \oG \nonumber \\[2pt]
&& +i\olambda_{\mu} [\Dc_{+\mu},\rho]
+ i\lambda_{\mu} [\Dc_{+\mu},\orho] 
+\epsilon_{\mu\nu\rho} \lambda_{\mu}
[\Dc_{-\nu},\olambda_{\rho}] \biggr]. \label{SYMaction_prototype}
\end{eqnarray}
The exact form w.r.t. all the supercharges of twisted $N=4\ D=3$
manifestly ensures the invariance of the action under any of the
SUSY transformation,
\begin{eqnarray}
s_{A} S^{N=4\ D=3}_{TSYM} =0,
\hspace{30pt}
s_{A}=(s,\os_{\mu},s_{\mu},\os).
\end{eqnarray}
By substituting $\Dc_{\pm\mu}=\partial_{\mu}-i(A_{\mu}\pm \phi^{(\mu)})$,
the action can be written as,
\begin{eqnarray} 
S^{N=4\ D=3}_{TSYM}&=& \int d^{3}x\ \mathrm{tr}\ \biggl[
\frac{1}{2}  
F_{\mu\nu}F_{\mu\nu}
-[\Dc_{\mu},\phi^{(\nu)}][\Dc_{\mu},\phi^{(\nu)}]
-\frac{1}{2}[\phi^{(\mu)},\phi^{(\nu)}][\phi^{(\mu)},\phi^{(\nu)}] 
+K^{2} + G\oG \nonumber \\[6pt]
&&+i\olambda_{\mu}[\Dc_{\mu},\rho] + i\lambda_{\mu}[\Dc_{\mu},\orho]
+\epsilon_{\mu\nu\rho}\lambda_{\mu}[\Dc_{\nu},\olambda_{\rho}]\nonumber \\[4pt]
&& +\olambda_{\mu}[\phi^{(\mu)},\rho] + \lambda_{\mu}[\phi^{(\mu)},\orho]
+i\epsilon_{\mu\nu\rho}\lambda_{\mu}[\phi^{(\nu)},\olambda_{\rho}]
\biggr],\label{N=4D=3SYMaction}
\end{eqnarray}
where $F_{\mu\nu}\equiv i[\Dc_{\mu},\Dc_{\nu}]$ are representing 
the field strength with the gauge covariant derivatives 
$\Dc_{\mu} \equiv \partial_{\mu} -iA_{\mu}$.
The tr is representing the trace for the gauge group.
One could notice that the kinetic terms, the potential terms
and the Yukawa coupling terms are naturally arising from the
combinations of the $\Dc_{\pm\mu}$ in (\ref{SYMaction_prototype}).
The action (\ref{N=4D=3SYMaction}) is the
continuum counterpart of the twisted $N=4\ D=3$ lattice SYM action 
proposed in \cite{DKKN3}. 

Here we have a couple of remarks regarding the action (\ref{N=4D=3SYMaction}).
First,  the action 
can be expressed
by re-writing the fermion kinetic terms and the Yukawa coupling terms 
in the following way,
\begin{eqnarray}
S^{N=4\ D=3}_{TSYM}&=& \int d^{3}x\ \mathrm{tr}\ \biggl[
\frac{1}{2}  
F_{\mu\nu}F_{\mu\nu}
-[\Dc_{\mu},\phi^{(\nu)}][\Dc_{\mu},\phi^{(\nu)}]
-\frac{1}{2}[\phi^{(\mu)},\phi^{(\nu)}][\phi^{(\mu)},\phi^{(\nu)}] 
+K^{2} + G\oG \nonumber \\[2pt]
&&+i\overline{\psi}_{i\alpha}(\gamma_{\mu})_{\alpha\beta}[\Dc_{\mu},\psi_{\beta i}] 
+\overline{\psi}_{i\alpha}[\phi^{(\mu)},\psi_{\alpha j}](\gamma_{\mu})_{ji} 
\biggr],
\label{N=4D=3SYMaction2}
\end{eqnarray}
where we introduced the ``untwisted" basis of fermions
$\psi_{\alpha i}$ and $\overline{\psi}_{i\alpha}$ 
defined by the expansions w.r.t. 
the $N=4\ D=3$ twisted fermions
$(\rho,\lambda_{\mu})$ and $(\orho,\olambda_{\mu})$, respectively,
\begin{eqnarray}
\psi_{\alpha i}&=& 
\frac{1}{\sqrt{3}}({\mathbf{1}}\rho +\gamma_{\mu}\lambda_{\mu})_{\alpha i},
\hspace{30pt}
\overline{\psi}_{i \alpha} \ =\ 
\frac{1}{\sqrt{3}}({\mathbf{1}}\orho +\gamma_{\mu}\olambda_{\mu})_{i \alpha},
\label{expansions}
\end{eqnarray}
The ${\bf 1}$ again denotes a two-by-two unit matrix
while the gamma matrices $\gamma_{\mu}$ are taken as the Pauli matrices,
$\gamma_{\mu}(\mu =1,2,3) = (\sigma_{1},\sigma_{2},\sigma_{3})$.
In terms of the untwisted fermions (\ref{expansions}), one could see 
the manifest invariance of the action (\ref{N=4D=3SYMaction2})
under the independent $SU(2)_{E}$ and $SU(2)_{R}$ rotations. 
The expressions (\ref{expansions})
also imply that the $N=4\ D=3$ twisted fermions 
have one-to-one correspondences with
a three dimensional Dirac-K\"ahler fermion components \cite{BJ}.
This is more clearly seen in the corresponding
lattice formulation given in \cite{DKKN3},
where the $N=4\ D=3$ twisted fermions $(\rho,\olambda_{\mu},\lambda_{\mu},\orho)$
are essentially 
embedded in the three dimensional lattice as a (0-form, 1-form, 2-form, 3-form),
respectively.

One should also notice from the second equality in (\ref{SYMaction_1stline})
that the $N=4\ D=3$ twisted SYM action can be expressed 
by the $s\overline{s}$-exact form,
\begin{eqnarray}
S^{N=4\ D=3}_{TSYM} &=& \frac{1}{2} s \overline{s}
\int d^{3}x \ \os_{3} s_{3} \ \mathrm{tr}\  \Dc_{-3}\Dc_{-3} \\[2pt]
&=& \frac{1}{2} s \overline{s}
\int d^{3}x \ \mathrm{tr}\ 
\biggl[
-\frac{1}{3}\epsilon_{\mu\nu\rho} \Dc_{-\mu}[\Dc_{-\nu},\Dc_{-\rho}]
+2\rho\overline{\rho}
\biggr]. 
\end{eqnarray}
After using the cyclic trace property, 
partial integrations and
the commuting nature of the derivative operators,
$[\partial_{\mu},\partial_{\nu}]=0$,
one finds that the above action 
can be written down as the $s\overline{s}$-exact form
on the following Chern-Simons type action attached with
the fermion bilinear term $\rho\overline{\rho}$,
\begin{eqnarray}
S^{N=4\ D=3}_{TSYM} &=&
\frac{1}{2} s\overline{s}\ S_{SCS}, \label{exact1} \\[2pt]
S_{SCS} &\equiv& 
\int d^{3}x \ \mathrm{tr}\ 
\biggl[
\epsilon_{\mu\nu\rho}(A^{-}_{\mu}[\partial_{\nu},A^{-}_{\rho}]
-\frac{i}{3}A^{-}_{\mu}[A^{-}_{\nu},A^{-}_{\rho}]) 
+ 2\rho\overline{\rho}
\biggr], 
\end{eqnarray}
where the symbols $A^{-}_{\mu}$ are defined by 
$A^{-}_{\mu}\equiv A_{\mu}-\phi^{(\mu)}$. 
It is rather striking to recognize that
the SYM action and the above type of Chern-Simons action are
intrinsically related by the twisted SUSY transformations.
This result is actually consistent with the (on-shell) formulations
of $N=4\ D=3$ twisted SYM with the auxiliary one-forms 
$B$ and $\overline{B}$ given in \cite{BT,GM}.  
Furthermore, thanks to the manifestly off-shell formulation here,
it is easy to show that the $S_{SCS}$ can also be expressed by the 
$\overline{s}_{1}s_{1}$ and $\overline{s}_{2}s_{2}$ exact forms,
\begin{eqnarray}
S_{SCS} = \int d^{3}x\ \os_{3}s_{3}\
\mathrm{tr}\ \Dc_{-3}\Dc_{-3}  =
\int d^{3}x\ \os_{1}s_{1}\ 
\mathrm{tr}\ \Dc_{-1}\Dc_{-1}  =
\int d^{3}x\ \os_{2}s_{2}\ 
\mathrm{tr}\ \Dc_{-2}\Dc_{-2}, \qquad
\end{eqnarray}
from which 
it obeys the invariance of the
$S_{SCS}$ under the six of the twisted SUSY transformations
$(s_{1},s_{2},s_{3},\os_{1},\os_{2},\os_{3})$,
\begin{eqnarray}
s_{\mu} S_{SCS} &= &
\os_{\mu} S_{SCS} \ =\ 0, \hspace{20pt}
(\mu = 1,2,3).
\end{eqnarray}
It is important to recognize here that the supercharges ($s_{\mu}$, $\os_{\mu}$)
and the sub-multiplet $(\Dc_{-\mu},\rho,\orho)$
form an off-shell closed sub-algebra
embedded in the entire $N=4\ D=3$ twisted SUSY algebra
(\ref{N=4D=3SYMalgebra1})-(\ref{N=4D=3SYMalgebra4}),
\begin{eqnarray}
\{s_{\mu},\os_{\nu}\}\varphi_{sub} 
&=& -\epsilon_{\mu\nu\rho}[\Dc_{-\rho},\varphi_{sub}], \\[2pt]
\{s_{\mu},s_{\nu}\}\varphi_{sub} &=& \{\os_{\mu},\os_{\nu}\}\varphi_{sub}
\ =\ 0,
\end{eqnarray}
where $\varphi_{sub}$ is representing any component of
the sub-multiplet $\varphi_{sub}=(\Dc_{-\mu},\rho,\orho)$.

Another important observation is that
the Lagrangian density in the $S_{SCS}$
transforms as a scalar only under the twisted rotations
$J^{diag}_{\mu}=J_{\mu}+R_{\mu}$ and not under the $J_{\mu}$ 
and the $R_{\mu}$ independently,
which implies that the action $S_{SCS}$ is purely 
a twisted object. 
As a consequence, one cannot perform ``untwisting" the action $S_{SCS}$
even in the flat spacetime. 
This should be compared to the rotational property of
$S^{N=4\ D=3}_{TSYM}$ which is invariant under
the $J_{\mu}$ and $R_{\mu}$ rotations independently.
The other type of $J^{diag}_{\mu}$ scalar Chern-Simons
type action can be found by noticing 
that the $S^{N=4\ D=3}_{TSYM}$
is also expressed as
the following form,
\begin{eqnarray}
S^{N=4\ D=3}_{TSYM} &=& \frac{1}{6}\int d^{3}x\
\os_{\mu}s_{\mu} s\os
\ \mathrm{tr}\ \Dc_{-\nu}\Dc_{-\nu} \\[2pt]
&\equiv& \frac{1}{2}s_{\mu}\overline{s}_{\mu}\ S'_{SCS}, 
\label{exact2}
\end{eqnarray}  
where $\mu$ and $\nu$ are summed up from $1$ to $3$.
The  
$S'_{SCS}$ is given by the Chern-Simons type action with
a mixed combination of $A^{\pm}_{\mu}\equiv A_{\mu}\pm \phi^{(\mu)}$ 
attached with another fermion bilinear terms $\lambda_{\mu}\olambda_{\mu}$, 
\begin{eqnarray}
S'_{SCS} &=& -\frac{1}{3}\int d^{3}x\  s\overline{s}
\ \mathrm{tr}\ \Dc_{-\nu}\Dc_{-\nu} \\[2pt]
&=& \int d^{3}x\  \mathrm{tr} 
\biggl[
-\frac{1}{3}\epsilon_{\mu\nu\rho}\Dc_{-\mu}[\Dc_{+\nu},\Dc_{+\rho}]
+\frac{2}{3}\lambda_{\mu}\overline{\lambda}_{\mu}
\biggr] \\[2pt]
&=& \int d^{3}x\ \mathrm{tr} \
\biggl[
\epsilon_{\mu\nu\rho}
\{
\frac{1}{3}(A^{+}_{\mu}+2A^{-}_{\mu})[\partial_{\nu},A^{+}_{\rho}]
-\frac{i}{3}A^{-}_{\mu}[A^{+}_{\nu},A^{+}_{\rho}]
\}
+\frac{2}{3}\lambda_{\mu}\overline{\lambda}_{\mu}
\biggr],
\end{eqnarray}
It is clearly seen from its exact form 
that the $S'_{SCS}$ is invariant under the
scalar type twisted SUSY transformations $s$ and $\os$,
\begin{eqnarray}
s S'_{SCS} &=& \overline{s} S'_{SCS} \ =\ 0.
\end{eqnarray} 
The above SUSY invariance is again supported by the existence of the
sub-algebra and the sub-multiplet 
embedded in the entire twisted $N=4\ D=3$ algebra 
(\ref{N=4D=3SYMalgebra1})-(\ref{N=4D=3SYMalgebra4}),
\begin{eqnarray}
\{s,\os \}\varphi'_{sub} &=& s^{2}\varphi'_{sub} \ =\ \os^{2}\varphi'_{sub} \ =\ 0,
\end{eqnarray}
which holds off-shell
for the sub-multiplet $\varphi'_{sub}=(\Dc_{\pm\mu},\lambda_{\mu},\olambda_{\mu})$.

Remembering that the $N=4\ D=3$ twisted SYM action
$S^{N=4\ D=3}_{TSYM}$ can be realized on the lattice
consistently with the lattice Leibniz rule conditions \cite{DKKN3},
one may wonder the above exactness
relation 
between the SYM and the super Chern-Simons
would be playing a key role also in realizing the
Chern-Simons on the lattice. 
Since addressing this topic is beyond the initial scope of this paper,
we keep this subject as our future study. 
In the next section we will shed light on the rather different 
aspect of the $N=4\ D=3$ twisted SYM, namely its dimensional
reduction aspect from $N=2\ D=4$ twisted SYM.

\section{Two possible twists of $N=4\ D=3$ SYM}
\label{two_possible_twists}

In this section, by a dimensional reduction of the $N=2\ D=4$ twisted SYM constraints,
we explore the two possible twists of $N=4\ D=3$ SYM
and we show that the $N=4\ D=3$ twisted SYM in the last section 
is essentially classified as the B-type twisted SYM.

The formulations of the $N=2\ D=4$ twisted SYM in terms of the superconnection
method are given in \cite{KKM,Alvarez_Labastida}. 
In \cite{KKM}, the detailed analysis originated from the $N=D=4$ Dirac-K\"ahler
point of view is also explicitly elaborated.
We start from the following $N=2\ D=4$ twisted SYM constraints 
in the twisted superspace $(x, \theta^{+},\theta^{+}_{\mu},\theta^{+}_{\rho\sigma})$
\cite{KKM},
\begin{eqnarray}
\{\D^{+},\D^{+}_{\mu}\} &=& -i\D_{\underline{\mu}},\hspace{30pt}
\{\D^{+}_{\rho\sigma},\D^{+}_{\mu}\} \ =\ +i\delta^{+}_{\rho\sigma\mu\nu}
\D_{\underline{\nu}}, \label{N=2D=4const1} \\
\{\D^{+},\D^{+}\} &=& -iW,  \hspace{30pt}
\{\D^{+}_{\mu\nu},\D^{+}_{\rho\sigma}\} \ =\ -i\delta^{+}_{\mu\nu\rho\sigma}W,
\label{N=2D=4const2}\\
\{\D^{+}_{\mu},\D^{+}_{\nu}\} &=& -i\delta_{\mu\nu}F, \hspace{30pt} 
\{ others\} \ =\ 0, \label{N=2D=4const3}
\end{eqnarray}
where the symbol $\delta^{+}_{\mu\nu\rho\sigma}$ 
is defined as
$\delta^{+}_{\mu\nu\rho\sigma}
\equiv \delta_{\mu\rho}\delta_{\nu\sigma}-\delta_{\mu\sigma}\delta_{\nu\rho}
+\epsilon_{\mu\nu\rho\sigma}$. 
The second rank tensor $\D^{+}_{\mu\nu}$ satisfies
the self-duality condition,
$\frac{1}{2}\epsilon_{\mu\nu\rho\sigma}\D^{+}_{\rho\sigma} = \D^{+}_{\mu\nu}$.
The symbols $(\D^{+},\D^{+}_{\mu},\D^{+}_{\rho\sigma})$
are denoting 
the supergauge covariant derivatives 
which consist of the $N=2\ D=4$ superderivatives 
$(D^{+},D^{+}_{\mu},D^{+}_{\rho\sigma})$ and the superconnections
$(\Gamma^{+},\Gamma^{+}_{\mu},\Gamma^{+}_{\rho\sigma})$,
\begin{eqnarray}
\D^{+} &=& D^{+} -i\Gamma^{+}
(\theta^{+},\theta^{+}_{\alpha},\theta^{+}_{\gamma\delta}), \\
\D^{+}_{\mu} &=& D^{+}_{\mu} -i\Gamma^{+}_{\mu}
(\theta^{+},\theta^{+}_{\alpha},\theta^{+}_{\gamma\delta}), \\
\D^{+}_{\rho\sigma} &=& D^{+}_{\rho\sigma} -i\Gamma^{+}_{\rho\sigma}
(\theta^{+},\theta^{+}_{\alpha},\theta^{+}_{\gamma\delta}), 
\end{eqnarray}
where the superderivatives $(D^{+},D^{+}_{\mu},D^{+}_{\rho\sigma})$ satisfy
the following $N=2\ D=4$ twisted SUSY algebra,
\begin{eqnarray}
\{D^{+},D^{+}_{\mu}\} &=& -i\partial_{\mu}, \hspace{20pt}
\{D^{+}_{\rho\sigma},D^{+}_{\mu}\} \ =\ +i\delta^{+}_{\rho\sigma\mu\nu}\partial_{\nu},
\hspace{20pt}
\{others\} \ =\ 0.
\end{eqnarray}
We denote the expansions of
the bosonic gauge covariant superfields $(\D^{+}_{\underline{\mu}},W,F)$
in the r.h.s. of (\ref{N=2D=4const1})-(\ref{N=2D=4const3}) as,
\begin{eqnarray}
\D_{\underline{\mu}} &=& \partial_{\mu}-i A_{\mu} + \cdots, \hspace{20pt}
W \ =\ A +\cdots, \hspace{20pt}
F \ =\ B +\cdots, \label{expansion}
\end{eqnarray}
where the $A_{\mu}$, 
$A$ and $B$ are representing the four dimensional gauge field and
the two independent scalar fields, respectively.
All the component fields in the $\D_{\underline{\mu}}$, $W$ and $F$
can be expressed by the combinations of the component fields 
embedded in the superconnections\
$(\Gamma^{+},\Gamma^{+}_{\mu},\Gamma^{+}_{\rho\sigma})$
subject to the constraints (\ref{N=2D=4const1})-(\ref{N=2D=4const3}). 
The dots in (\ref{expansion}) denote the possible 
$\theta^{+}_{A}=(\theta^{+},\theta^{+}_{\mu},\theta^{+}_{\rho\sigma})$ 
expansion terms.

The $SO(4)_{E}\times SU(2)_{R}$
rotational properties of the (super)covariant
derivatives are given by \cite{KKM}, 
\begin{eqnarray}
[J^{+}_{\mu\nu},\D^{+}] &=& +\frac{i}{2}\D^{+}_{\mu\nu}, \hspace{48pt}
[J^{-}_{\mu\nu},\D^{+}_{\rho}]  \ =\  -\frac{i}{2}
\delta^{-}_{\mu\nu\rho\sigma}\D^{+}_{\sigma},\\[0pt]
[J^{+}_{\mu\nu},\D^{+}_{\rho\sigma}]&=&-\frac{i}{2}\delta^{+}_{\mu\nu\rho\sigma}\D^{+}
+\frac{i}{4}(\delta^{+}_{\mu\nu\rho\lambda}\D^{+}_{\sigma\lambda}
-\delta^{+}_{\mu\nu\sigma\lambda}\D^{+}_{\rho\lambda}),\\[4pt]
[J^{-}_{\mu\nu},\D^{+}] &=& [J^{+}_{\mu\nu},\D^{+}_{\rho}] 
=[J^{-}_{\mu\nu},\D^{+}_{\rho\sigma}] = 0,\\[2pt]
[J^{+}_{\mu\nu},\D_{\underline{\rho}}]
&=&-\frac{i}{2}\delta^{+}_{\mu\nu\rho\sigma}\D_{\underline{\sigma}},
\hspace{30pt}
[J^{-}_{\mu\nu},\D_{\underline{\rho}}]
\ =\ -\frac{i}{2}\delta^{-}_{\mu\nu\rho\sigma}\D_{\underline{\sigma}},\\[2pt]
[J^{+}_{\mu\nu},W] &=&[J^{-}_{\mu\nu},W] = [J^{+}_{\mu\nu},F] 
= [J^{-}_{\mu\nu},F] =0, 
\end{eqnarray}
\begin{eqnarray}
[R^{+}_{\mu\nu},\D^{+}] &=& -\frac{i}{2}\D^{+}_{\mu\nu}, \hspace{48pt} 
[R^{+}_{\mu\nu},\D^{+}_{\rho}]  \ =\  -\frac{i}{2}
\delta^{-}_{\mu\nu\rho\sigma}\D^{+}_{\sigma},\\[0pt]
[R^{+}_{\mu\nu},\D^{+}_{\rho\sigma}]&=&+\frac{i}{2}\delta^{+}_{\mu\nu\rho\sigma}\D^{+}
+\frac{i}{4}(\delta^{+}_{\mu\nu\rho\lambda}\D^{+}_{\sigma\lambda}
-\delta^{+}_{\mu\nu\sigma\lambda}\D^{+}_{\rho\lambda}),\\[4pt]
[R^{-}_{\mu\nu},\D^{+}] &=& [R^{-}_{\mu\nu},\D^{+}_{\rho}] 
= [R^{-}_{\mu\nu},\D^{+}_{\rho\sigma}]= 0,\\[8pt]
[R^{+}_{\mu\nu},\D_{\underline{\rho}}]&=&[R^{+}_{\mu\nu},W] 
= [R^{+}_{\mu\nu},F] = 0, 
\end{eqnarray}
where $J^{+}_{\mu\nu}$ and $J^{-}_{\mu\nu}$ denote
the self-dual and anti-selfdual part of 
$SO(4)_{E}$ Euclidean Lorentz generators
while the $R^{+}_{\mu\nu}$ 
denote $SU(2)_{R}$ internal rotation generators.
Note that $R^{+}_{\mu\nu}$ is also subject to the self-duality condition,
$R^{+}_{\mu\nu}=\frac{1}{2}\epsilon_{\mu\nu\rho\sigma}R^{+}_{\rho\sigma}$.
The symbols $\delta^{\pm}_{\mu\nu\rho\sigma}
\equiv \delta_{\mu\rho}\delta_{\nu\sigma}-\delta_{\mu\sigma}\delta_{\nu\rho}
\pm \epsilon_{\mu\nu\rho\sigma}$  
are projecting the self-dual and anti-selfdual part, respectively.

Now we perform the dimensional reduction of the $N=2\ D=4$ twisted SYM constraints
to the $N=4\ D=3$ constraints. 
We take the component fields independent of the fourth direction $x_{4}$
and denote the (super)gauge covariant derivatives 
in terms of the following $N=4\ D=3$ notations,
\begin{eqnarray}
&&\D^{+} \rightarrow \D^{1}, \hspace{20pt}
\D^{+}_{1} \rightarrow \D^{2}_{1}, \hspace{20pt}
\D^{+}_{2} \rightarrow \D^{2}_{2}, \hspace{20pt}
\D^{+}_{3} \rightarrow \D^{2}_{3}, \hspace{20pt}
\D^{+}_{4} \rightarrow -\D^{2}, \qquad \label{D=3notation1} \\[2pt]
&&\D^{+}_{12}\rightarrow \D^{1}_{3}, \hspace{16pt}
\D^{+}_{13} \rightarrow -\D^{1}_{2}, \hspace{9pt}
\D^{+}_{14} \rightarrow \D^{1}_{1}, \hspace{20pt}
\D^{+}_{\underline{4}} \rightarrow -G, \label{D=3notation2}
\end{eqnarray}
where the $G$ is representing the scalar field originated from the
gauge field in fourth dimension.
By the dimensional reduction,
the original Euclidean rotational group $SO(4)_{E}$
is reduced into $SU(2)_{E}$ which is the covering group of
three dimensional Euclidean rotation $SO(3)_{E}$,
while the original internal symmetry $SU(2)_{R}$ remains intact.
Furthermore, as is pointed out in \cite{SW},
we have yet another $SU(2)$ symmetry associated with the
$N=4\ D=3$ SUSY algebra, which is denoted as $SU(2)_{N}$. 
The existence of two independent internal symmetries
$SU(2)_{R}$ and $SU(2)_{N}$ leads to the two possible
topological twists \cite{BT,GM} which we will explicitly see in the following.

In terms of three dimensional notation (\ref{D=3notation1})-(\ref{D=3notation2}), 
the constraints (\ref{N=2D=4const1})-(\ref{N=2D=4const3})
turn into the 
following form of $N=4\ D=3$ SYM constraints,
\begin{eqnarray}
\{\D^{a},\D^{b}_{\mu}\} &=& -i \epsilon^{ab}\D_{\underline{\mu}}, \hspace{30pt}
\{\D^{a}_{\mu},\D^{b}_{\nu}\} \ =\ 
+i \epsilon^{ab}\epsilon_{\mu\nu\rho}\D_{\underline{\rho}}
-i\delta_{\mu\nu}\phi^{ab}, \label{N=4D=3constA1} \\
\{\D^{a},\D^{b}\} &=& -i\phi^{ab}, \hspace{47pt}
\{others\} \ =\ 0, \label{N=4D=3constA2}
\end{eqnarray}
where the subscripts $\mu,\nu,\rho$ run from $1$ to $3$
while the $SU(2)_{N}$ superscripts $a,b$ take $1$ or $2$.
The scalar fields $\phi^{ab}$ form the triplet state of $SU(2)_{N}$   
where each component of $\phi^{ab}$ is defined by
the scalar fields introduced in (\ref{N=2D=4const2}), (\ref{N=2D=4const3}) and
(\ref{D=3notation2}), 
\begin{eqnarray}
\phi^{11} \ =\ W, \hspace{20pt}
\phi^{12} &=& \phi^{21} \ =\ G, \hspace{20pt}
\phi^{22} \ =\ F.
\end{eqnarray}
The transformations of the supercovariant derivatives
under the whole symmetry group $SU(2)_{E}\times SU(2)_{R}\times SU(2)_{N}$ 
are given by 
\begin{eqnarray}
[J_{\mu},\D^{a}] &=& +\frac{i}{2}\D^{a}_{\mu}, 
\hspace{30pt}
[J_{\mu},\D^{a}_{\nu}] \ =\ -\frac{i}{2}\epsilon_{\mu\nu\rho}\D^{a}_{\rho}
-\frac{i}{2}\delta_{\mu\nu}\D^{a}, \label{JA1} \\[2pt]
[R_{\mu},\D^{a}] &=& -\frac{i}{2}\D^{a}_{\mu}, 
\hspace{30pt}
[R_{\mu},\D^{a}_{\nu}] \ =\ -\frac{i}{2}\epsilon_{\mu\nu\rho}\D^{a}_{\rho}
+\frac{i}{2}\delta_{\mu\nu}\D^{a}, \label{JA2} \\[2pt]
[N_{\mu},\D^{a}_{A}] &=& \frac{1}{2}(\gamma_{\mu})^{ab}\D^{a}_{A}, \label{JA3}
\end{eqnarray}
where $J_{\mu}$, $R_{\mu}$ and $N_{\mu}$ denote
the generators of $SU(2)_{E}$, $SU(2)_{R}$ and $SU(2)_{N}$, respectively.
\footnote{The $J_{\mu}$ and $R_{\mu}$ are defined by 
$J_{1}=J_{14}^{+}+J_{14}^{-}$,\ $J_{2}=-J_{13}^{+}-J_{13}^{-}$,\ 
$J_{3}=J_{12}^{+}+J_{12}^{-}$,\
$R_{1}=R_{14}^{+}+R_{14}^{-}$,\ $R_{2}=-R_{13}^{+}-R_{13}^{-}$ and 
$R_{3}=R_{12}^{+}+R_{12}^{-}$ in terms of $N=2\ D=4$ notation.}
In the last line, $\D^{a}_{A}$ represents
any of $(\D^{a},\D^{a}_{\mu})$.
The gamma matrices $\gamma_{\mu}$ are taken as the Pauli matrices,
$\gamma_{\mu}(\mu=1,2,3)=(\sigma_{1},\sigma_{2},\sigma_{3})$. 
The generators $J_{\mu}$, $R_{\mu}$ and $N_{\mu}$ obey
the independent $SU(2)$ algebra,
\begin{eqnarray}
[J_{\mu},J_{\nu}] &=& -i\epsilon_{\mu\nu\rho}J_{\rho},\hspace{10pt}
[R_{\mu},R_{\nu}] \ =\ -i\epsilon_{\mu\nu\rho}R_{\rho},\hspace{10pt}
[N_{\mu},N_{\nu}] \ =\ -i\epsilon_{\mu\nu\rho}N_{\rho}, \\[2pt]
[J_{\mu},R_{\nu}] &=& [J_{\mu},N_{\nu}] \ =\ [R_{\mu},N_{\nu}] \ =\ 0.
\end{eqnarray}

The first topological twist of $N=4\ D=3$, 
which is called the A-type or the super BF type twist in the literatures,
is given by taking the diagonal subgroup of $SU(2)_{E}\times SU(2)_{R}$
\cite{BT,GM}.
After the twisting, the entire rotational symmetries are governed by the
twisted Lorentz rotation $(SU(2)_{E}\times SU(2)_{R})_{diag}$
generated by $J^{diag}_{\mu}\equiv J_{\mu}+R_{\mu}$
and the internal rotation $SU(2)_{N}$ by $N_{\mu}$,
\begin{eqnarray}
[J^{diag}_{\mu},\D^{a}] &=& 0, 
\hspace{30pt}
[J^{diag}_{\mu},\D^{a}_{\nu}] \ =\ -i\epsilon_{\mu\nu\rho}\D^{a}_{\rho} 
\hspace{30pt}
[N_{\mu},\D^{a}_{A}] \ =\ \frac{1}{2}(\gamma_{\mu})^{ab}\D^{a}_{A},
\end{eqnarray}
namely, $\D^{a}$ and $\D^{a}_{\mu}$ are transforming as,
\begin{eqnarray}
\D^{a}:({\bf{1}},{\bf{2}}), \hspace{30pt}
\D^{a}_{\mu}:({\bf{3}},{\bf{2}}),
\end{eqnarray} 
of $(SU(2)_{E}\times SU(2)_{R})_{diag} \times SU(2)_{N}$.
It is also easy to see from the constraints (\ref{N=4D=3constA1})
and (\ref{N=4D=3constA2})
that the gauge field $A_{\mu}$
and the scalar fields are transforming as,
\begin{eqnarray}
A_{\mu}:({\bf{3}},{\bf{1}}), \hspace{30pt}
\phi^{ab}:({\bf{1}},{\bf{3}}). \label{A-type_A_phi}
\end{eqnarray} 
Notice that the $\phi^{ab}$ still transform
as scalars after the A-type twist.
By analyzing the Jacobi identities together with
the constraints (\ref{N=4D=3constA1})
and (\ref{N=4D=3constA2}), one can construct the corresponding
off-shell SYM multiplet which consists of 
the gauge field $A_{\mu}$, the 
scalar fields $\phi^{ab}$
as well as
the twisted fermions
$\rho^{a},\lambda_{\mu}^{a}$ and the bosonic auxiliary field $H_{\mu}$
transforming as,
\begin{eqnarray}
\rho^{a}:({\bf{1}},{\bf{2}}), \hspace{30pt}
\lambda^{a}_{\mu}:({\bf{3}},{\bf{2}}), \hspace{30pt}
H_{\mu}:({\bf{3}},{\bf{1}}).
\end{eqnarray} 
The SUSY transformations of the component fields 
and the corresponding SYM action of the A-type twist
can also be obtained through the Jacobi 
identity analyses under the constraints 
(\ref{N=4D=3constA1})-(\ref{N=4D=3constA2})
just as in the previous section.

The second topological twist of $N=4\ D=3$, 
which is called the B-type or the Blau-Thompson type twist,
is given by taking the diagonal subgroup of $SU(2)_{E}\times SU(2)_{N}$
\cite{BT,GM}.
One sees that the basis of $\D^{a}_{A}$
appeared in (\ref{N=4D=3constA1}) and (\ref{N=4D=3constA2}) is 
not appropriate for the B-type twist
since the operations of $J_{\mu}$ and $N_{\mu}$ in (\ref{JA1}) and (\ref{JA3})
are not on the same footing.
One of the appropriate basis for the B-type twist could be found
after taking the following linear combinations
of the super gauge covariant derivatives,
\begin{eqnarray}
\D'^{1} &\equiv& +\frac{1}{2}(\D^{1}+i\D^{1}_{1}+\D^{2}_{2}+i\D^{2}_{3}),
\hspace{20pt}
\D'^{2} \ \equiv\ +\frac{1}{2}(\D^{2}-i\D^{2}_{1}-\D^{1}_{2}+i\D^{1}_{3}), 
\label{recombi1} \quad  \\
\D'^{1}_{1} &\equiv& -\frac{i}{2}(\D^{1}+i\D^{1}_{1}-\D^{2}_{2}-i\D^{2}_{3}),
\hspace{20pt}
\D'^{2}_{1} \ \equiv\ +\frac{i}{2}(\D^{2}-i\D^{2}_{1}+\D^{1}_{2}-i\D^{1}_{3}), 
\label{recombi2}  \\
\D'^{1}_{2} &\equiv& -\frac{1}{2}(\D^{2}+i\D^{2}_{1}-\D^{1}_{2}-i\D^{1}_{3}),
\hspace{20pt}
\D'^{2}_{2} \ \equiv\  +\frac{1}{2}(\D^{1}-i\D^{1}_{1}+\D^{2}_{2}-i\D^{2}_{3}), 
\label{recombi3}  \\
\D'^{1}_{3} &\equiv& -\frac{i}{2}(\D^{2}+i\D^{2}_{1}+\D^{1}_{2}+i\D^{1}_{3}),
\hspace{20pt}
\D'^{2}_{3} \ \equiv\ -\frac{i}{2}(\D^{1}-i\D^{1}_{1}-\D^{2}_{2}+i\D^{2}_{3}), 
\label{recombi4} 
\end{eqnarray} 
and making the slight re-definitions of $R_{\mu}$ and $N_{\mu}$,
\begin{eqnarray}
R'_{1} &\equiv & -R_{3}, \hspace{20pt}
R'_{2} \ \equiv\ -R_{2}, \hspace{20pt}
R'_{3} \ \equiv\ -R_{1}, \\
N'_{1} &\equiv& -N_{3}, \hspace{20pt}
N'_{2} \ \equiv\ -N_{2}, \hspace{20pt}
N'_{3} \ \equiv\ -N_{1}.
\end{eqnarray} 
We then see that the relations (\ref{N=4D=3constA1})-(\ref{N=4D=3constA2})
are re-expressed as
\begin{eqnarray}
\{\D'^{a},\D'^{b}_{\mu}\} &=& -i\epsilon^{ab}(\D_{\underline{\mu}}+\phi^{(\mu)}), 
\label{N=4D=3constB1} \\
\{\D'^{a}_{\mu},\D'^{b}_{\nu}\}  
& =& +i\epsilon^{ab}\epsilon_{\mu\nu\rho}(\D_{\underline{\rho}}-\phi^{(\rho)}), 
\hspace{20pt}
\{others\} \ =\ 0, \label{N=4D=3constB2}
\end{eqnarray}
where the $\phi^{(\mu)}$ are given in terms of $W$, $F$ and $G$, 
\begin{eqnarray}
\phi^{(1)}\equiv -iG, \hspace{20pt}
\phi^{(2)}\equiv +\frac{1}{2}(W+F),\hspace{20pt}
\phi^{(3)}\equiv -\frac{i}{2}(W-F).
\end{eqnarray}
Accordingly, the relations (\ref{JA1})-(\ref{JA3}) are also re-expressed by
\begin{eqnarray}
[J_{\mu},\D'^{a}] &=& +\frac{i}{2}\D'^{a}_{\mu}, 
\hspace{30pt}
[J_{\mu},\D'^{a}_{\nu}] \ =\ -\frac{i}{2}\epsilon_{\mu\nu\rho}\D'^{a}_{\rho}
-\frac{i}{2}\delta_{\mu\nu}\D'^{a}, \label{JB1} \\[2pt]
[N'_{\mu},\D'^{a}] &=& -\frac{i}{2}\D'^{a}_{\mu}, 
\hspace{30pt}
[N'_{\mu},\D'^{a}_{\nu}] \ =\ -\frac{i}{2}\epsilon_{\mu\nu\rho}\D'^{a}_{\rho}
+\frac{i}{2}\delta_{\mu\nu}\D'^{a}, \label{JB2} \\[2pt]
[R'_{\mu},\D'^{a}_{A}] &=& \frac{1}{2}(\gamma_{\mu})^{ab}\D'^{a}_{A}. \label{JB3}
\end{eqnarray}
Notice that, after the re-definitions (\ref{recombi1})-(\ref{recombi4}),
the $SU(2)_{E}$ and the $SU(2)_{N}$ are operating on the same footing
as if the role of the $SU(2)_{N}$ and the $SU(2)_{R}$ were interchanged.
The second topological twist can be appropriately performed 
on this basis by taking the diagonal subgroup of $SU(2)_{E}\times SU(2)_{N}$.
After the B-type twist, the rotational symmetries are governed by
the generators
$J'^{diag}_{\mu}\equiv J_{\mu}+N'_{\mu}$ and $R'_{\mu}$ which are representing 
the twisted Lorentz of 
$(SU(2)_{E}\times SU(2)_{N})_{diag}$
and the internal rotation generators of $SU(2)_{R}$, respectively.
From (\ref{JB1})-(\ref{JB3}), one could obviously see 
\begin{eqnarray}
[J'^{diag}_{\mu},\D'^{a}] &=& 0, 
\hspace{20pt}
[J'^{diag}_{\mu},\D'^{a}_{\nu}] \ =\ -i\epsilon_{\mu\nu\rho}\D'^{a}_{\rho}, 
\hspace{20pt}
[R'_{\mu},\D'^{a}_{A}] \ =\ \frac{1}{2}(\gamma_{\mu})^{ab}\D'^{a}_{A}.
\end{eqnarray}
Namely the $\D'^{a}$ and $\D'^{a}_{\mu}$ are transforming as,
\begin{eqnarray}
\D'^{a}:({\bf{1}},{\bf{2}}), \hspace{30pt}
\D'^{a}_{\mu}:({\bf{3}},{\bf{2}}), \label{B-type_nablas}
\end{eqnarray} 
of $(SU(2)_{E}\times SU(2)_{N})_{diag} \times SU(2)_{R}$.
The gauge fields $A_{\mu}$ and the scalar fields $\phi^{(\mu)}$
are accordingly transforming as
\begin{eqnarray}
A_{\mu}:({\bf{3}},{\bf{1}}), \hspace{30pt}
\phi^{(\mu)}:({\bf{3}},{\bf{1}}). \label{B-type_bosons}
\end{eqnarray} 
Notice that, after the B-twist, the scalar fields $\phi^{(\mu)}$
transform as a three dimensional vector just like the gauge fields.
The B-type twisted fermions $(\rho^{a},\lambda_{\mu}^{a})$ 
are transforming in a similar way as the A-type twisted ones
while the auxiliary field $H^{ab}$ transform as a $SU(2)_{R}$
triplet states,
\begin{eqnarray}
\rho^{a}:({\bf{1}},{\bf{2}}), \hspace{30pt}
\lambda^{a}_{\mu}:({\bf{3}},{\bf{2}}), \hspace{30pt}
H^{ab}:({\bf{1}},{\bf{3}}). \label{B-type_fermions}
\end{eqnarray} 

It comes clear from the constraints (\ref{N=4D=3constB1})-(\ref{N=4D=3constB2})
and the representations of the components
(\ref{B-type_nablas})-(\ref{B-type_fermions}) 
that the $N=4\ D=3$ twisted SYM in Sec. \ref{offshellSYM}
can be essentially identified as the B-type twisted SYM
described in this section.
We have the following notational identifications
for the $SU(2)_{R}$ doublet and triplet states,
\begin{eqnarray}
(\D_{A}, \overline{\D}_{A}) \leftrightarrow \D'^{a}_{A}, \hspace{30pt}
(\rho,\overline{\rho}) \leftrightarrow \rho^{a},\hspace{30pt}
(\lambda_{\mu},\overline{\lambda}_{\mu})
\leftrightarrow \lambda^{a}_{\mu},
\hspace{30pt}
(G,\overline{G},K) \leftrightarrow H^{ab}, 
\end{eqnarray}
where the appropriate sign re-definitions are understood.
Note that the internal symmetry generators $R_{\mu}$ 
in Sec. \ref{offshellSYM} can be identified as 
$N'_{\mu}$ in this section.
The above correspondences indicate that
the lattice formulation of $N=4\ D=3$ twisted SYM
given in \cite{DKKN3} is essentially classified as the B-type twisted SYM.

\section{The twisted $N=2\ D=4$ and  $N=4\ D=3$ SYM from the lattice point of view}
\label{lattice_realization_other_types}

Knowing the classification of 
the two inequivalent topological twists of $N=4\ D=3$ SYM
and remembering the lattice realization of B-type twisted SYM \cite{DKKN3},
one should  ask the possibility of formulating the A-type twisted SYM
on the lattice as well.
In this section, we consider the lattice Leibniz rule
and the lattice gauge covariance
for the $N=2\ D=4$ and $N=4\ D=3$ twisted SUSY algebra and multiplet.
We then explicitly see that
the B-type SYM is only the case
which can be consistently realized on the three dimensional lattice
satisfying these criteria.

\begin{figure}
\begin{center}
\begin{minipage}{60mm}
\begin{center}
\includegraphics[width=50mm]{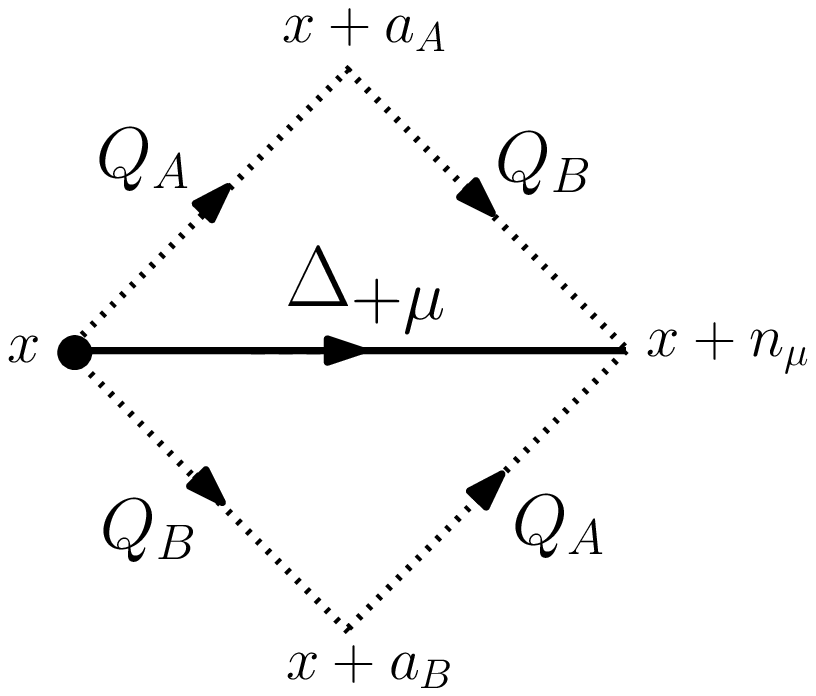}
\caption{Lattice SUSY algebra subject to the condition (\ref{alg_forward})}
\label{s_Delta1}
\end{center}
\end{minipage}
\hspace{20pt}
\begin{minipage}{60mm}
\begin{center}
\includegraphics[width=50mm]{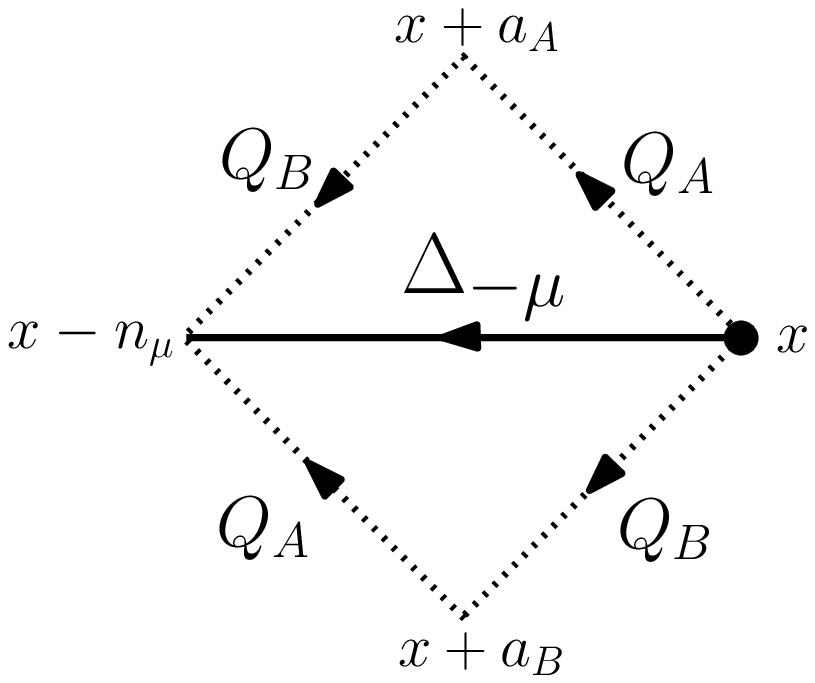}
\caption{Lattice SUSY algebra subject to the condition (\ref{alg_backward})}
\label{s_Delta2}
\end{center}
\end{minipage}
\end{center}
\end{figure}

Let us briefly remind the basic idea of  
the lattice Leibniz rule 
introduced in \cite{DKKN1,DKKN2}.
The importance of Leibniz rule is also stressed in the context of 
the non-commutative differential geometry on the lattice \cite{KK}.
Since on the lattice there are no infinitesimal translations, 
the derivative operators should be replaced by
the corresponding difference operators of either forward  or backward,
$P_{\mu}=i\partial_{\mu} \rightarrow i\Delta_{\pm\mu}$.
The operation of the difference operators on the lattice is naturally 
 defined by the following type of ``shifted" commutators,
\begin{eqnarray}
(\Delta_{\pm\mu}\Phi(x)) &=& \Delta_{\pm\mu}\Phi(x)-\Phi(x\pm n_{\mu})\Delta_{\pm\mu},
\end{eqnarray}
where
the $n_{\mu}$ denote three dimensional lattice unit vectors. 
The $\Delta_{\pm \mu}$ 
are located on links from $x$ to $x\pm n_{\mu}$, respectively,
and taking the unit values for the generic site $x$,
\begin{eqnarray}
\Delta_{\pm\mu}=(\Delta_{\pm\mu})_{x\pm n_{\mu},x} = \mp 1.
\end{eqnarray}
Correspondingly, we define the lattice supercharges $Q_{A}$ on the links
from $x$ to $x+a_{A}$ whose operations are defined by the ``shifted"
(anti)commutators,
\begin{eqnarray}
(Q_{A}\Phi(x)) = (Q_{A})_{x+a_{A},x}\Phi(x)-(-)^{|\Phi|}
\Phi(x+a_{A})(Q_{A})_{x+a_{A},x},
\end{eqnarray}
where the symbol $|\Phi|$ takes the value of 0 or 1 for the 
bosonic or the fermionic $\Phi$, respectively.
Since the supercharges $Q_{A}$ are located on links, the anti-commutators
of supercharges are naturally defined by the successive connections
of link supercharges,
\begin{eqnarray}
\{Q_{A},Q_{B}\}_{x+a_{A}+a_{B},x} =
(Q_{A})_{x+a_{A}+a_{B},x+a_{B}}(Q_{B})_{x+a_{B},x}
+(Q_{B})_{x+a_{A}+a_{B},x+a_{A}}(Q_{A})_{x+a_{A},x}. 
\label{link_anti-comm}
\end{eqnarray}
In terms of these ingredients, the lattice SUSY algebra can be expressed as
\begin{eqnarray}
\{Q_{A},Q_{B}\}_{x+a_{A}+a_{B},x} = (\Delta_{\pm\mu})_{x,x\pm n_{\mu}}
\end{eqnarray}
provided the following lattice Leibniz rule conditions hold 
(see Fig. \ref{s_Delta1} and Fig. \ref{s_Delta2}), 
\begin{eqnarray}
a_{A}+a_{B}&=& +n_{\mu}\ \ \ {\rm for}\ \ \ \Delta_{+\mu}, \label{alg_forward}\\
a_{A}+a_{B}&=& -n_{\mu}\ \ \ {\rm for}\ \ \ \Delta_{-\mu}. \label{alg_backward}
\end{eqnarray}
It has been pointed out that the Dirac-K\"ahler twisted $N=D=2$,
$N=4\ D=3$ and $D=N=4$ SUSY algebra can satisfy such conditions 
\cite{DKKN1,DKKN2,DKKN3}. 
Since these successful examples are explained in the references in detail,
it is rather instructive here to begin with the Leibniz rule conditions for the 
twisted $N=2\ D=4$ SUSY algebra and then see how the situations are improved
by the dimensional reduction to the  twisted $N=4\ D=3$.

The twisted $N=2\ D=4$ SUSY algebra is given by
\begin{eqnarray}
\{Q^{+},Q^{+}_{\mu}\} &=& i\partial_{\mu}, \hspace{20pt}
\{Q^{+}_{\rho\sigma},Q^{+}_{\mu}\} \ =\ -i\delta^{+}_{\rho\sigma\mu\nu}\partial_{\nu},
\hspace{20pt}
\{others\} \ =\ 0, \label{N=2D=4Qalg}
\end{eqnarray}
where $\mu,\nu,\rho,\sigma$ run from $1$ to $4$. 
The symbols $(Q^{+},Q^{+}_{\mu},Q^{+}_{\mu\nu})$ denote
the $N=2\ D=4$ twisted supercharges which transform as
a scalar, a vector and a second rank self-dual tensor, respectively.
The projector $\delta^{+}_{\mu\nu\rho\sigma}
=\delta_{\mu\rho}\delta_{\nu\sigma}-\delta_{\mu\sigma}\delta_{\nu\rho}
+\epsilon_{\mu\nu\rho\sigma}$ picks up only the self-dual part.
The lattice Leibniz rule conditions associated with 
the lattice counterpart of (\ref{N=2D=4Qalg}) are expressed as,
for example, 
\begin{eqnarray}
a^{+}+a_{1}^{+} &=& \pm n_{1}, \hspace{20pt} 
a_{12}^{+} + a_{1}^{+} \ =\ \pm n_{2}, \ \ \ {\rm etc}.,
\end{eqnarray}
where the signs are chosen to be positive (negative) if the
corresponding difference operator is the forward (backward) type.  
All the conditions associated with the twisted $N=2\ D=4$ algebra 
are summarized in Table \ref{N=2D=4Leibniz_table}.
Viewing these conditions, one can easily notice that
they are actually over-constrained.
For example, the total sum of $a_{A}$ along the diagonal part of
Table \ref{N=2D=4Leibniz_table} gives,
\begin{eqnarray}
\sum a_{A} &=& (a^{+} + a^{+}_{1}) + (a^{+}_{12} + a^{+}_{2}) +
(a^{+}_{13} + a^{+}_{3}) + (a^{+}_{14} + a^{+}_{4}) \nonumber \\
&=& \pm n_{1} \pm n_{1} \pm n_{1} \pm n_{1} \ \propto\  n_{1}\ \ \mathrm{or}\ \ 0, 
\label{diag}
\end{eqnarray}
while one of the off-diagonal combinations give rise to the
different value of the total sum,
\begin{eqnarray}
\sum a_{A} &=& (a^{+}_{13} + a^{+}_{1}) + (a^{+}_{12} + a^{+}_{2}) +
(a^{+}_{14} + a^{+}_{3}) + (a^{+} + a^{+}_{4}) \nonumber \\
&=& \pm n_{3} \pm n_{1} \pm n_{2} \pm n_{4},
\label{off-diag}
\end{eqnarray}
which indicates that the $a_{A}$ cannot have any definite values.
The lattice Leibniz rule conditions for twisted $N=2\ D=4$ 
thus do not have any consistent solutions.
\begin{table}
\renewcommand{\arraystretch}{1.4}
\renewcommand{\tabcolsep}{10pt}
\begin{center}
\begin{minipage}{70mm}
\begin{center}
\begin{tabular}{c|cccc}
& $a^{+}_{1}$ & $a^{+}_{2}$ & $a^{+}_{3}$ & $a^{+}_{4}$ \\ \hline
$a^{+}$ & $\pm n_{1}$ & $\pm n_{2}$ & $\pm n_{3}$ & $\pm n_{4}$  \\ 
$a^{+}_{12}$ & $\pm n_{2}$ & $\pm n_{1}$ & $\pm n_{4}$ & $\pm n_{3}$  \\ 
$a^{+}_{13}$ & $\pm n_{3}$ & $\pm n_{4}$ & $\pm n_{1}$ & $\pm n_{2}$  \\ 
$a^{+}_{14}$ & $\pm n_{4}$ & $\pm n_{3}$ & $\pm n_{2}$ & $\pm n_{1}$  
\end{tabular}
\caption{Leibniz rule conditions for the $N=2\ D=4$ twisted SUSY algebra}
\label{N=2D=4Leibniz_table}
\end{center}
\end{minipage}
\hspace{20pt}
\begin{minipage}{70mm}
\begin{center}
\begin{tabular}{c|cccc}
& $a^{2}_{1}$ & $a^{2}_{2}$ & $a^{2}_{3}$ & $a^{2}$ \\ \hline
$a^{1}$ & $\pm n_{1}$ & $\pm n_{2}$ & $\pm n_{3}$ &   \\ 
$a^{1}_{3}$ & $\pm n_{2}$ & $\pm n_{1}$ &  & $\pm n_{3}$  \\ 
$a^{1}_{2}$ & $\pm n_{3}$ &  & $\pm n_{1}$ & $\pm n_{2}$  \\ 
$a^{1}_{1}$ &  & $\pm n_{3}$ & $\pm n_{2}$ & $\pm n_{1}$  
\end{tabular}
\caption{Leibniz rule conditions after the dimensional reduction}
\label{Dim_Red_Leibniz_table}
\end{center}
\end{minipage}
\end{center}
\end{table}

Now we perform a dimensional reduction to three dimensions
by truncating the fourth dimension.
By employing the $SU(2)_{R}\times SU(2)_{N}$ manifestly covariant notation
in the last section, the $N=4\ D=3$ twisted SUSY algebra 
of either the A-type or the B-type can be written as
\begin{eqnarray}
\{Q^{a},Q^{b}_{\mu}\} &=& +i\epsilon^{ab}\partial_{\mu}, \hspace{30pt}
\{Q^{a}_{\mu},Q^{b}_{\nu}\} \ =\ -i\epsilon^{ab}\epsilon_{\mu\nu\rho}\partial_{\rho}.
\label{3Dalg}
\end{eqnarray}
The Leibniz rule conditions for the lattice counterpart of (\ref{3Dalg})
is summarized in Table \ref{Dim_Red_Leibniz_table}.
Notice that the off-diagonal summations such as (\ref{off-diag})
turn to be irrelevant after the dimensional reduction 
since there are no conditions arising from the fourth direction.
Taking a look at the total sum of $a_{A}$'s in terms of 
the relevant combinations, 
one could realize that the total sum of $a_{A}$'s should vanish
in order for the $N=4\ D=3$ Leibniz rule conditions
to be satisfied,
which means that one should have two forward difference 
and two backward difference operators 
for each direction.
One of the possible choices for the $N=4\ D=3$ lattice SUSY algebra
is thus given by, for example,
\begin{eqnarray}
\{Q^{a},Q^{b}_{\mu}\} &=& +i\epsilon^{ab}\Delta_{+\mu}, \hspace{30pt}
\{Q^{a}_{\mu},Q^{b}_{\nu}\} \ =\ -i\epsilon^{ab}\epsilon_{\mu\nu\rho}\Delta_{-\rho},
\label{3Dalg_lat}
\end{eqnarray}
which is associated with the Leibniz rule condition,
\begin{eqnarray}
a^{a}+a^{b}_{\mu} &=& +|\epsilon^{ab}|n_{\mu}, \hspace{30pt}
a^{a}_{\mu}+a^{b}_{\nu} \ =\ -|\epsilon^{ab}||\epsilon_{\mu\nu\rho}|n_{\rho}.
\label{3DLeibniz_cond}
\end{eqnarray}
As is presented in \cite{DKKN3}, 
the consistent solutions for the conditions (\ref{3DLeibniz_cond})
are given by
\begin{eqnarray}
a^{1} &=& (arbitrary), \ \ \ \ \ a^{2}_{\mu} \ =\ +n_{\mu} -a^{1},
\label{solution1} \\
a^{1}_{\mu} &=& -\sum_{\lambda\neq \mu}n_{\lambda}+a^{1}, \ \ \ 
a^{2} \ =\ +\sum_{\lambda=1}^{3}n_{\lambda} -a^{1}. \label{solution2}
\end{eqnarray}

As is also stressed in \cite{DKKN3}, the eight supercharges of $N=4\ D=3$
twisted SUSY algebra have one-to-one correspondences with all the possible
simplicial elements in three dimensions, namely, $0$-form, $1$-form, $2$-form
and $3$-form whose total number of components is $1+3+3+1=8$.
This geometrical consistency with the Dirac-K\"ahler picture of the fermions
essentially provides the reason why the $N=4\ D=3$ 
twisted algebra can be exactly realized on the three dimensional lattice.
From this viewpoint, there is no wonder why the $N=2\ D=4$ lattice Leibniz rule
conditions do not have any consistent solutions,
because in four dimensions we have $0$-form, $1$-form, $2$-form,
$3$-form and $4$-form whose total number of components is $1+4+6+4+1=16$.
We obviously need the Dirac-K\"ahler twisted $N=D=4$ SUSY algebra
with sixteen supercharges 
to be exactly realized on the four dimensional lattice,
as is already 
pointed out in \cite{DKKN2}. 

Let us now turn to address 
the possibility of formulating the A-type twisted SYM on the lattice.
We first remind that, in the manifestly gauge covariant formulation of
the lattice SYM \cite{DKKN2,DKKN3},
each supercharge $(Q_{A})_{x+a_{A},x}$ is replaced by the corresponding 
fermionic gauge link variable $(\D_{A})_{x+a_{A},x}$ 
whose gauge variation is given by
\begin{eqnarray}
(\D_{A})_{x+a_{A},x} &\rightarrow& G_{x+a_{A}}(\D_{A})_{x+a_{A},x}G^{-1}_{x},
\end{eqnarray} 
where $G_{x}$ denotes the finite gauge transformation at the site $x$. 
In order for the SYM multiplet to be realized on the lattice,
we need to take care of not only the Leibniz rule itself but also the
lattice gauge covariance of the entire SYM multiplet as well.
Although there is no distinction between the A-type
and the B-type in the lattice realizations of
the SUSY algebra itself (\ref{3Dalg_lat})
except for interchanging the roles of the $SU(2)_{R}$ and the $SU(2)_{N}$,
the situation becomes quite different when one comes to 
the gauge covariance on the lattice.

For the B-type twisted SYM which is subject to
the constraints (\ref{N=4D=3constB1}) and (\ref{N=4D=3constB2}), 
we can successfully introduce the bosonic gauge link variables,
$(\U_{\pm\mu})_{x\pm n_{\mu},x}=(e^{\pm i(A_{\mu}\pm \phi^{(\mu)})})_{x\pm n_{\mu},x}$, 
as the lattice realization of the gauge covariant derivatives,
$\mp (\partial_{\mu}-i(A_{\mu}\pm \phi^{(\mu)}))$ \cite{DKKN3}.
Note that the scalar fields $\phi^{(\mu)}$, 
transforming as a three dimensional vector after the twisting,
are embedded in the bosonic gauge link variables $\U_{\pm\mu}$.
In terms of these link variables,
the lattice counterpart of the constraints (\ref{N=4D=3constB1})
and (\ref{N=4D=3constB2}) can be expressed as
\begin{eqnarray}
\{\D^{a},\D^{b}_{\mu}\}_{x+a^{a}+a^{b}_{\mu},x} 
&=& +i\epsilon^{ab}(\U_{+\mu})_{x+n_{\mu},x}, 
\label{N=4D=3const_latB1} \\
\{\D^{a}_{\mu},\D^{b}_{\nu}\}_{x+a^{a}_{\mu}+a^{b}_{\nu},x}  
& =& +i\epsilon^{ab}\epsilon_{\mu\nu\rho}(\U_{-\rho})_{x-n_{\rho},x}, 
\label{N=4D=3const_latB2}
\\[2pt]
\{others\} & =& 0, 
\label{N=4D=3const_latB3}
\end{eqnarray}
where the anti-commutators in the l.h.s. are defined as
the link anti-commutators as in the supercharge case (\ref{link_anti-comm}),
\begin{eqnarray}
\{\D^{a},\D^{b}_{\mu}\}_{x+a^{a}+a^{b}_{\mu},x} &\equiv&
(\D^{a})_{x+a^{a}+a^{b}_{\mu},x+a^{b}_{\mu}}
(\D^{b}_{\mu})_{x+a^{b}_{\mu},x}
+(\D^{b}_{\mu})_{x+a^{b}_{\mu}+a^{a},x+a^{a}}
(\D^{a})_{x+a^{a},x} \qquad
\end{eqnarray}
Notice that the Leibniz rule conditions (\ref{3DLeibniz_cond})
are nothing but 
the gauge covariance conditions for the B-type twisted SYM constraints
(\ref{N=4D=3const_latB1}) and (\ref{N=4D=3const_latB2}).
Furthermore, once the starting constraints are 
given in a gauge covariant manner,
all the analyses of the Jacobi identities
automatically respect the gauge covariance on the lattice
thanks to the link definition of 
the (anti-)commutators. 
In this way, the gauge covariance for the B-type twisted SYM
is manifestly maintained 
on the lattice.
\footnote{As is claimed in \cite{DKKN3},
we need to introduce covariantly constant fermionic parameters $\eta_{A}$
in order to maintain the gauge covariance
associated with the twisted SUSY variations 
of the component fields on the lattice.}

As for the A-type twisted multiplet in contrast,
one finds that the scalar fields $\phi^{ab}$
are embedded in the constraints (\ref{N=4D=3constA1}) and (\ref{N=4D=3constA2})
as follows,
\begin{eqnarray}
\{\D^{a},\D^{b}\}=
\{\D^{a}_{1},\D^{b}_{1}\}=
\{\D^{a}_{2},\D^{b}_{2}\}=
\{\D^{a}_{3},\D^{b}_{3}\}= -i\phi^{ab}, \label{phi_const}
\end{eqnarray}
This ``diagonal" embedding of the scalar fields is
directly related the fact that the $\phi^{ab}$
are transforming as scalars under $(SO(3)_{E}\times SO(3)_{R})_{diag}$
even after the twisting (See (\ref{A-type_A_phi})). 
The crucial observation here is that the equalities in (\ref{phi_const})
can never be simultaneously satisfied on the lattice  
since the solutions for the lattice Leibniz rule
(\ref{solution1})-(\ref{solution2}) indicate that
each anti-commutator should be located on a different link each other.
For example, in the case of $a\neq b$ we have
the four anti-commutators obviously located on different links each other,
\begin{eqnarray}
\{\D^{a},\D^{b}\}_{x+n_{1}+n_{2}+n_{3},x}, \ 
\{\D^{a}_{1},\D^{b}_{1}\}_{x+n_{1}-n_{2}-n_{3},x}, \  
\{\D^{a}_{2},\D^{b}_{2}\}_{x-n_{1}+n_{2}-n_{3},x}, \  
\{\D^{a}_{3},\D^{b}_{3}\}_{x-n_{1}-n_{2}+n_{3},x}. \ \nonumber \\
\end{eqnarray}
Likewise, we do not have any chance  to
simultaneously satisfy the equalities in (\ref{phi_const}) 
in the case of $a=b$, either.
Thus, the scalar fields $\phi^{ab}$ in the A-type twisted multiplet 
can never be located on any definite links on the three dimensional lattice. 
We conclude that, amongst two types of the $N=4\ D=3$ twisted SYM,
only the B-type twisted multiplet can be realized on the lattice keeping the
Leibniz rule and the gauge covariance on the lattice.
Namely, the formulation given in \cite{DKKN3}
is the unique lattice realization of $N=4\ D=3$ twisted SYM satisfying
these criteria.

One may wonder why the A-type multiplet fails on the lattice 
even though the starting constraints for the A-type
(\ref{N=4D=3constA1})-(\ref{N=4D=3constA2})
and the B-type (\ref{N=4D=3constB1})-(\ref{N=4D=3constB2})
are related each other by the linear combinations 
(\ref{recombi1})-(\ref{recombi4}).
One should remind here that although the A-type and B-type
basis are related each other, 
those two give rise to the inequivalent theories
after the twisting 
since they respect the different diagonal subgroups
of the Lorentz and the internal rotations.
The above analysis suggests that only the B-type twisted rotational subgroup
can be survived on the three dimensional lattice consistently with the
gauge covariance of the lattice SUSY multiplet.

It should be mentioned here that 
we had a similar situation also in the twisted $N=D=2$.
The existense of two inequivalent twists in the $N=D=2$ SYM
is originated from the two inequivalent internal symmetries, 
the ghost number $U(1)$ and the $SO(2)_{R}$ internal rotation.\footnote{The author
thanks I. Kanamori for his comments and discussions.}
In the lattice formulation proposed in \cite{DKKN2},  
we introduced the scalar fields in such a way that
they transform as a two dimensional vector after the twisting.
We could maintain the gauge covariance of the $N=D=2$ twisted SYM
multiplet on the lattice
by embedding the gauge fields and the scalar fields
in the bosonic gauge link variables
just as in the B-type twist of $N=4\ D=3$ explained above.
The other twisted basis of $N=D=2$ turned out not to
accommodate the lattice gauge covariance
just as in the case of the A-type twisted $N=4\ D=3$.  
The lattice formulation given in \cite{DKKN2}
is thus providing the unique lattice realization of $N=D=2$ twisted 
SYM compatible with the lattice Leibniz rule and the gauge covariance 
on the lattice.

\section{Summary \& Discussions}
An entirely off-shell formulation of the $N=4\ D=3$ twisted SYM is presented.
We employ the twisted superconnection method 
in order to provide an manifestly gauge covariant off-shell framework. 
Although the formulation is given in terms of the twisted basis,
the resulting SYM action respects the entire symmetry group of 
the three dimensional Lorentz rotations $SU(2)_{E}$ and 
the internal rotations $SU(2)_{R}$ in the flat spacetime.
We also explore the two inequivalent twisted SYM of $N=4\ D=3$
rather explicitly 
and we then show that the recent proposal of $N=4\ D=3$ twisted SYM
on the lattice \cite{DKKN3} is essentially classified as the B-type twisted SYM.
We also consider the possibility of realizing
the $N=2\ D=4$ twisted SYM as well as the
$N=4\ D=3$ A-type twisted SYM on the lattice
by analyzing the lattice Leibniz rule
and the gauge covariance on the lattice.
We  then show that the $N=2\ D=4$ twisted SUSY algebra
cannot satisfy the Leibniz rule conditions on the 
four dimensional lattice and 
that the A-type twisted SYM multiplet cannot be compatible
with the gauge covariance on the three dimensional lattice.
The analyses show that the lattice formulation
given in \cite{DKKN3} is the unique realization of 
$N=4\ D=3$ twisted SYM on the lattice
satisfying these criteria.
In the same respect, 
we also mentioned that the two dimensional lattice SYM formulation
given in \cite{DKKN1} is providing the unique formulation 
of $N=D=2$ twisted SYM on the lattice.

In this paper, we also explicitly derive the twisted SUSY exact relation between 
$N=4\ D=3$ twisted SYM
and the super Chern-Simons entirely in the off-shell regime.
Thanks to the off-shell structure, we clarify the 
twisted SUSY invariant nature of these super Chern-Simons actions.
We point out that the existence of the sub-algebra and the sub-multiplet
in the $N=4\ D=3$ is responsible for the twisted SUSY invariance.
We observe that these relations are also playing 
important roles when studying a possible formulation of 
the super Chern-Simons on the lattice. 
The result of this analysis will be given elsewhere.

\section*{Acknowledgments}

The author would like to thank 
S. Arianos,  F. Bruckmann, S. Catterall, A. D'Adda, I. Kanamori, J. Kato, 
N. Kawamoto, A. Miyake and J. Saito and T. Takimi 
for the discussions and comments.
The author is supported by Department of Energy US Government,
Grant No. FG02-91ER 40661.

\end{document}